\documentclass[a4paper,fleqn,usenatbib,useAMS]{mnras}

\usepackage[T1]{fontenc}
\usepackage{ae,aecompl}

\usepackage{siunitx}
\usepackage[T1]{fontenc}

\usepackage{multirow}
\usepackage{ae,aecompl}
\usepackage{footnote}
\usepackage{tikz}
\usepackage[normalem]{ulem}
\usepackage{caption}
\usepackage{subcaption}
\usepackage{graphicx}	
\usepackage{amsmath}	
\usepackage{amssymb}	

\title[Kinematical and chemical structures of dense cores ]{Studying the chemical and kinematical structures of dense cores TMC-1C, L1544, and TMC-1 in the Taurus molecular cloud using the CCS and NH$_{3}$ observations}

\author[Koley, A.]{
Atanu Koley $^{1, 2}$  \thanks{E-mail:atanuphysics15@gmail.com}
\\
$^{1}$Joint Astronomy Programme, Indian Institute of Science, Bangalore, 560012, India\\
$^{2}$Department of Physics,  Indian Institute of Science, Bangalore, 560012, India \\
}

\date{Accepted XXX. Received YYY; in original form ZZZ}

\pubyear{2015}

\begin{document}
\label{firstpage}
\pagerange{\pageref{firstpage}--\pageref{lastpage}}
\maketitle

\begin{abstract}
Measurement of chemical and kinematic structures in prestellar cores is essential for better understanding the star formation process. Here, we study the three prestellar cores (TMC-1C, L1544, and TMC-1) of the Taurus molecular cloud by means of the thioxoethenylidene (CCS) radical and ammonia (NH$_{3}$) molecule observed with Karl G. Jansky Very Large Array telescope in D, C, and CNB configurations. Our main results are based on the CCS observation of the TMC-1C core, showing complex structures are present. Spatial offset relative to dust emission is observed in the CCS radical. Across a wide region around the dust peak, inward motion is found through the CCS radical. We have calculated the infall velocity and measured the turbulence inside the core. The turbulence is found to be subsonic. We obtain the virial parameter $\alpha$ is $<$ 1. Thus, thermal and non-thermal motions cannot prevent the collapse. Spatial incoherence of the CCS and NH$_{3}$ is observed from the integrated intensity maps in these cores, suggesting that these molecules trace different environments of the cores.We compare the integrated flux densities of CCS with previous single-dish data and find that a small amount of flux is recovered in the interferometric observations, indicating the presence of significant diffuse emission in favorable conditions for producing CCS.

\end{abstract}

\begin{keywords} ISM: general - ISM: molecules - ISM: kinematics and dynamics, galaxies: star formation.


\end{keywords}


\section{Introduction}
Star formation is one of the prime events in the cosmos. Stars are born in dense cold environments called cloud cores. These dense cores are broadly classified into two categories: starless prestellar cores and star-containing protostellar cores \citep{2021A&A...653A..15N, 2007ApJ...671.1839S,2004A&A...416..191T,1992ApJ...392..551S}. Among those, the former one is of two types, young and evolved cores. Due to the absence of existing stars, starless cores have more privileged over the others for understanding the star formation process. To analyze different chemical and physical properties (e.g., density, temperature, turbulence, chemical abundance, magnetic field, etc.) in these cores, observations of molecular species and dust continuum mapping are useful. In particular, researchers have studied the abundances and spatial distributions of different molecules and dust, the strength and nature of the magnetic field and turbulence, etc. \citep{2019PASJ...71..117N,2018ApJ...864...82D,2008A&A...487..993K,2007ApJ...657..838S,2007ApJ...671.1839S,2007ApJ...663.1174S,2006A&A...450..569P,1998ApJ...504..207B,1981MNRAS.194..809L}. All these pieces of information are necessary for a complete understanding of the fragmentation process of dense cores. \\

H$_{2}$ molecule, the principal constituent in the cores, is mostly invisible. Therefore to study these cores, measurements of other molecules are required. Many observational studies exist in the literature on the pre-stellar cores using various molecules. However, as far as we are aware, a significant fraction of the observations were done with single-dish telescopes \citep{2021A&A...653A..15N,2007ApJ...657..838S,2006A&A...450..569P,1992ApJ...392..551S,1998ApJ...504..207B,2004A&A...416..191T}. These measurements could not properly study the chemical and physical structures into the small spatial scales of the cores or specifically inside the core nuclei. With this motivation, we have studied three prestellar cores (TMC-1C, L1544, and TMC-1) in Taurus molecular cloud with the  Very Large Array telescope (VLA) in different configurations to obtain all the cores' structures. For our analysis, we have used CCS radical and NH$_{3}$ molecule. One advantage of studying CCS  in conjunction with NH$_{3}$ molecule is that one of the molecules is in carbon-bearing categories, and another is in nitrogen-bearing categories. Astrochemical model \citep{2003ApJ...593..906A} tells us that carbon-bearing species (e.g., CS, CCS, etc.) are more abundant in the early stage of the cores and decrease along with the core evolution. Whereas nitrogen-bearing species (e.g., NH$_{3}$, NH$_{2}^{+}$, etc.) are more prominent in the evolved stage of the cores. Thus, the column density ratio and spatial distribution of CCS and NH$_{3}$ is an important parameter for studying the evolutionary stages of the cores \citep{1992ApJ...392..551S}. 
Despite the abundance ratio,  these molecules can also be an observational gauge to measure the magnetic field (through CCS) and kinetic temperature  (through NH$_{3}$)  in molecular clouds. Due to the low ionization fraction in the molecular cloud, Zeeman splitting and dust polarizations are the two main methods for studying the magnetic field \citep{2019PASJ...71..117N,2012SSRv..166..293H,1999sf99.proc..175S}. Magnetic field measurement from CCS radical will be discussed in another upcoming paper (Koley et al. 2022, in prep.). Regarding the turbulence, its strength and nature can be obtained after studying the kinetic temperature and all the nonthermal effects ( e.g., rotation, collapse, expansion, etc.) \citep{2007ApJ...657..838S,1998ApJ...504..207B,1981MNRAS.194..809L} from the spectra. If the dense cores are in the latter stage of their evolution and are on the verge of star formation, then spectral properties due to the collapse can also be studied through these molecules \citep{2012ApJ...759L..37C,2005ApJ...620..800D,1999ARA&A..37..311E,1996ApJ...465L.133M}. Overall full chemical and kinematical properties of the cores can be inferred from these spectral studies.\\
Earlier, \citet{2011ApJ...739L...4R} studied two out of these three aforementioned cores with CCS 2$_{1}$-1$_{0}$ transition (22.344 GHz) with VLA in DNC configuration. After the successful detection of their measurements, here, we extend the study with an additional NH$_{3}$ molecule. Our main results are based on the TMC-1C core studied with CCS 2$_{1}$-1$_{0}$ (22.344 GHz) transition observed with VLA in D configuration. In the first part, we will concentrate on this sensitive observation. After that, in the second part, we will briefly discuss the three cores, TMC-1C, L1544, and TMC-1, observed with high angular resolution observations (VLA in C and CNB configurations) with the  CCS 2$_{1}$-1$_{0}$ (22.344 GHz) and  NH$_{3}$(J,K = 1,1) (23.694 GHz) transitions.\\


In \S\ref{different cores}, we focus mainly on the Taurus molecular cloud and TMC-1C core. Other cores are briefly discussed in the later section. Observation and data analysis of VLA D configuration CCS emission of TMC-1C core are presented in \S\ref{observation and data analysis}. Our main chemical and kinematical results of the TMC-1C core based on the VLA  D configuration are presented in \S\ref{discussions}. In \S\ref{three cores}, we concentrate on the three cores, TMC-1C, L1544, and TMC-1, observed with the VLA in high-resolution observation (C and CNB). At last, we draw our main conclusions and future aims in \S\ref{conclusions}.\\

\begin{figure*}
\includegraphics[width=6.83in,height=5.3in,angle=0]{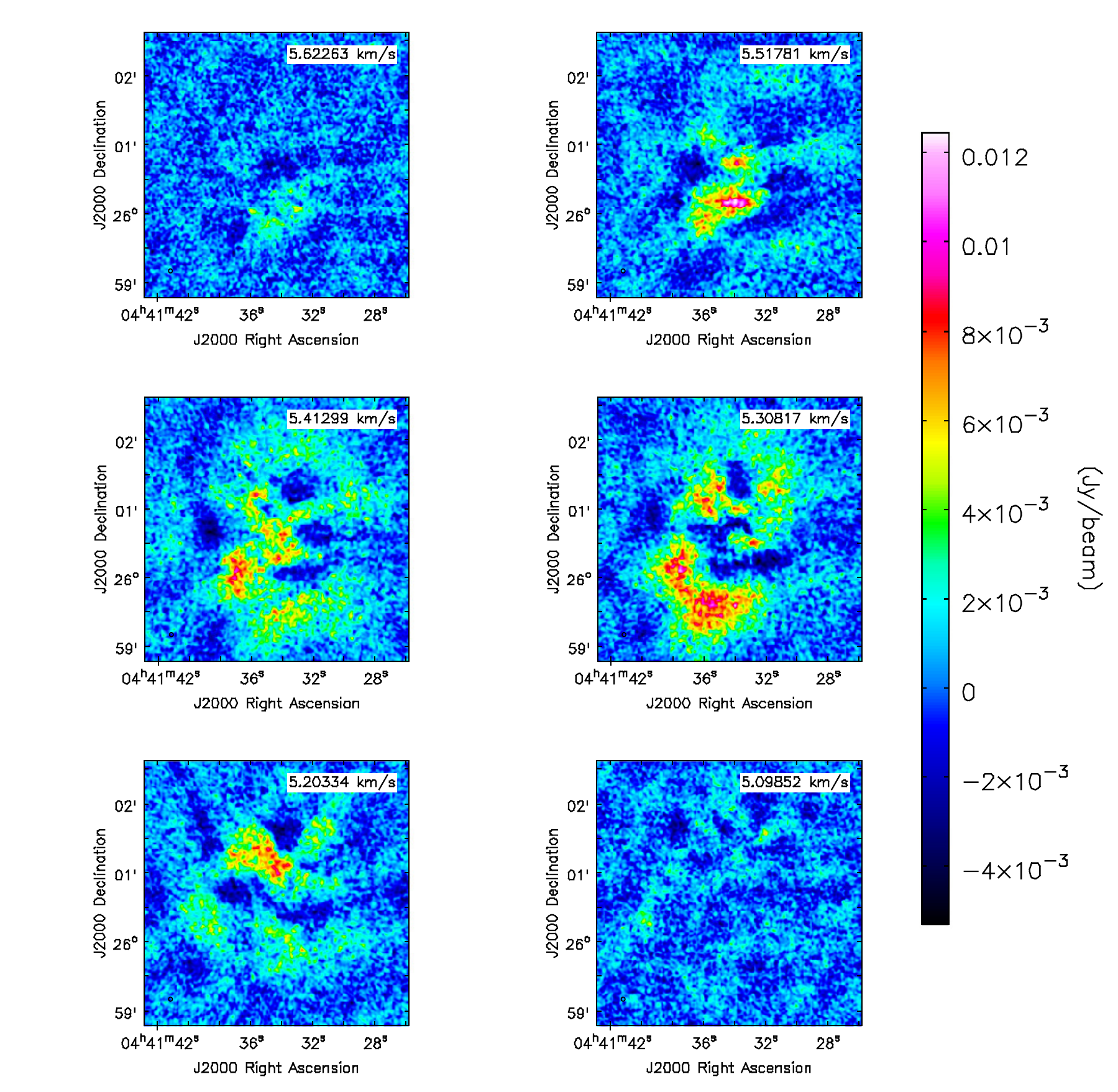}
 \caption{Channel maps of the CCS (2$_{1}$-1$_{0}$) 22.3 GHz line  observed with the VLA  in D configuration (before the primary beam correction).}
 \label{fig:fig1}
\end{figure*}

\begin{figure}
\includegraphics[width=3.4in,height=3.4in, angle=0]{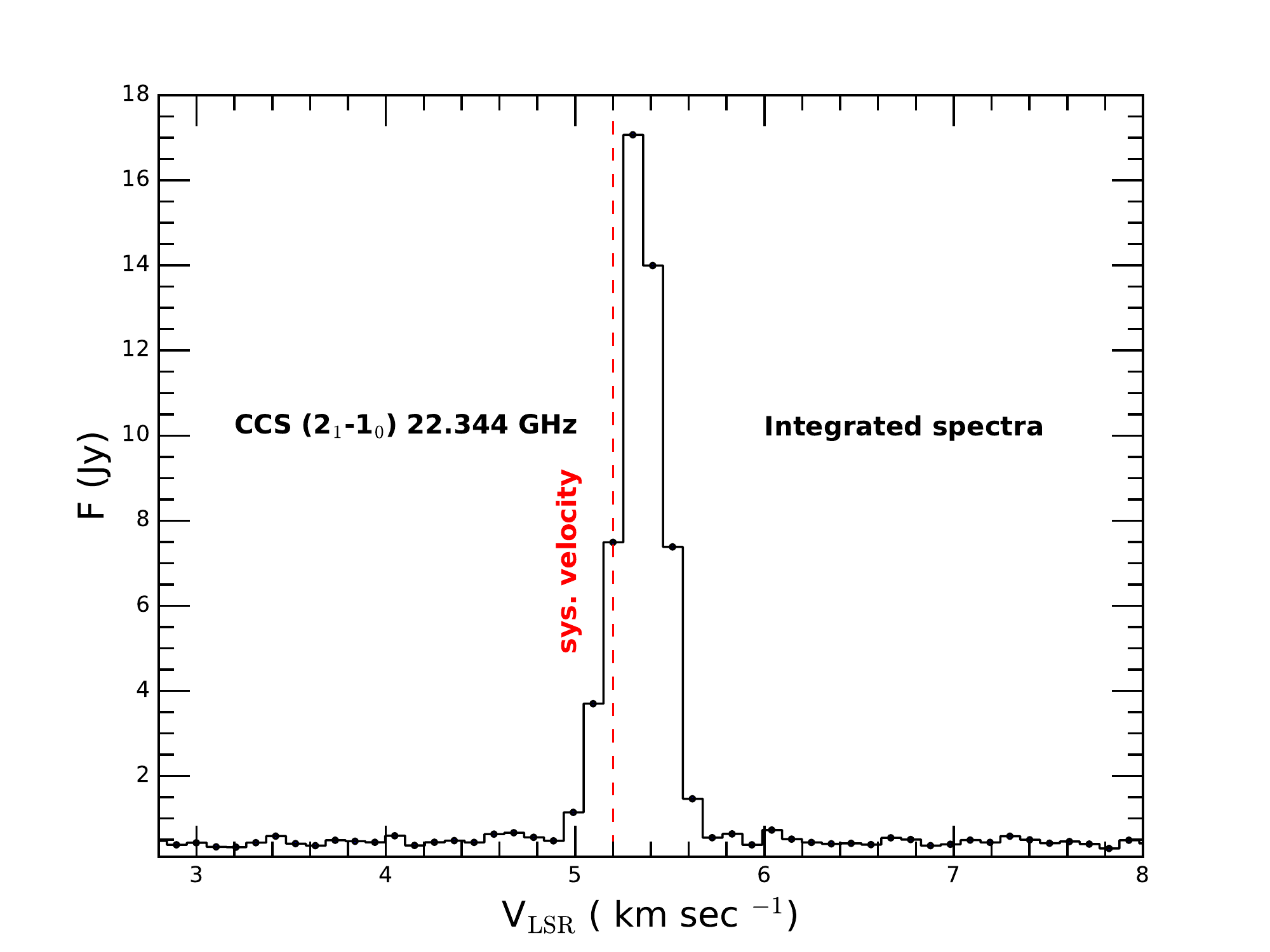}

  \caption{Integrated spectra of CCS (2$_{1}$-1$_{0}$) 22.3 GHz line towards the TMC-1C core.  RMS noise of the spectra is $\sim$ 0.23 Jy.}
 \label{fig:fig2}
\end{figure}

\begin{figure}
\includegraphics[width=3.5in,height=3.5in,angle=0]{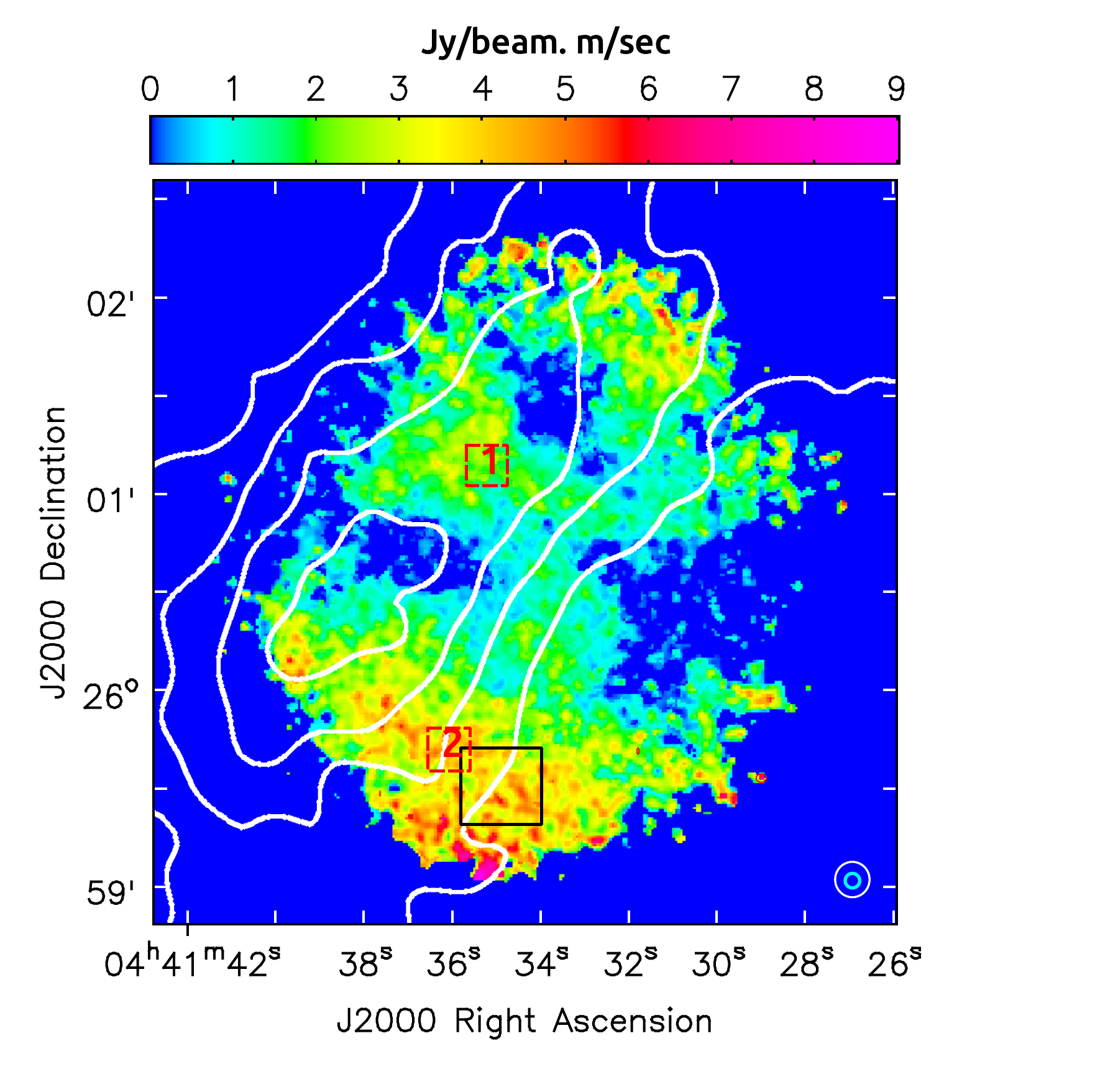}
 \caption{Overplot of integrated intensity maps of CCS (2$_{1}$-1$_{0}$) 22.3 GHz emission and dust 1200 $\mu$m emission. CCS emission is shown in color plot and dust emission is in contour plot. Contour levels are 15, 23, 30, 37 mJy/beam respectively. Black square box is the region where the magnetic field is tentatively detected (Koley et al. 2022, in prep.). Two red-squared boxes are the regions where the infall velocity has been calculated from the bule-skewed profile.  }
 \label{fig:fig3}
\end{figure}

\begin{figure*}
\includegraphics[width=7.0in,height=4.6in, angle=0]{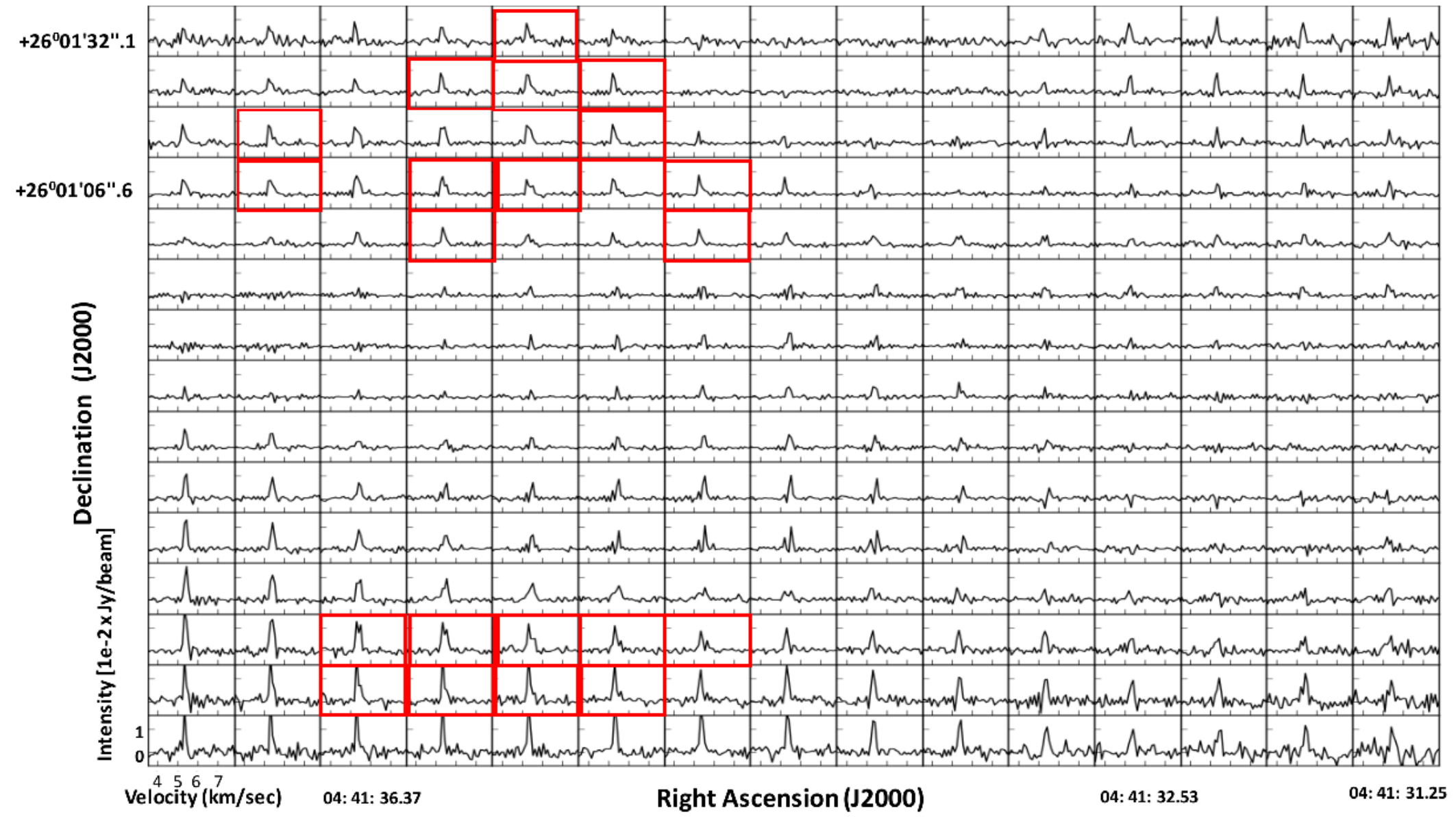}
  \caption{Spectra of CCS (2$_{1}$-1$_{0}$) 22.3 GHz line across the TMC-1C core after binning the whole area into 15 $\times$ 15 regions. Red boxes are the regions where the spectra are consistent with inward motions.}
 \label{fig:fig4}
\end{figure*}

\begin{figure*}
\includegraphics[width=7.0in,height=4.4in,angle=0]{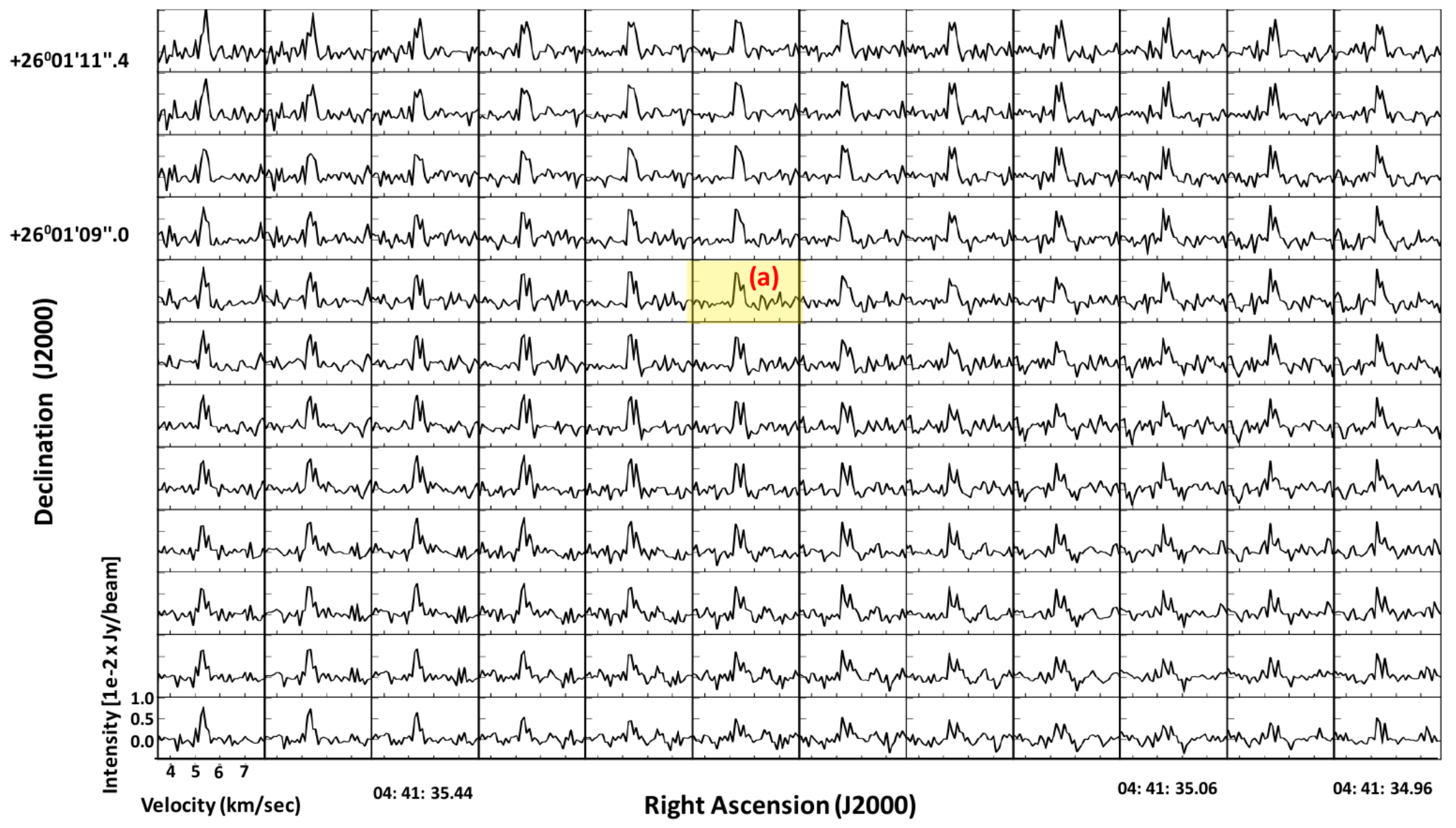}\\
\includegraphics[width=7.0in,height=4.4in,angle=0]{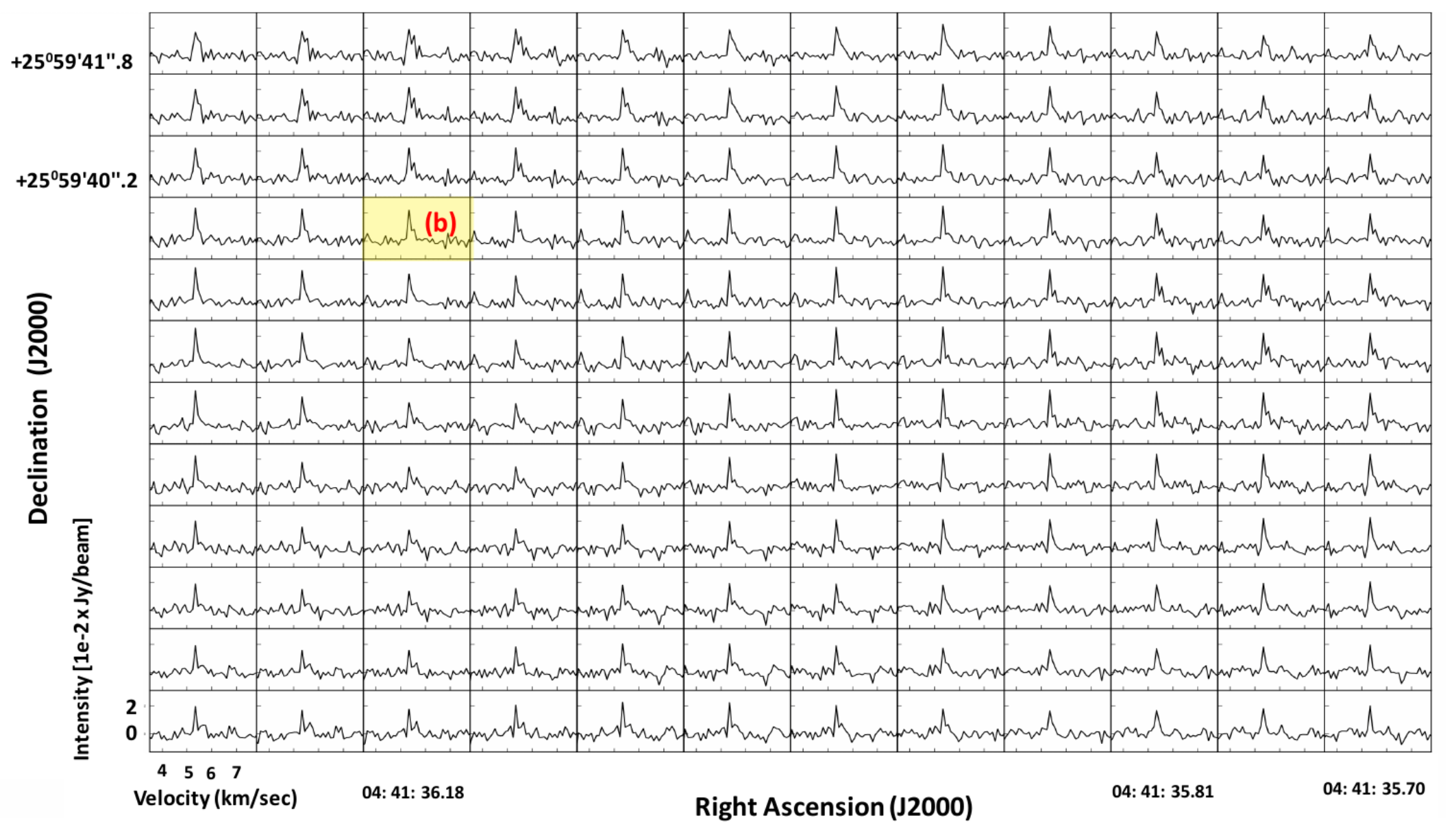}\\

  \caption{Upper Panel: Pixel by pixel spectra towards the region marked by the symbol 1 in the figure \ref{fig:fig3}. The highlighted yellow region marked by the symbol \textbf{a} is the pixel where infall velocity is calculated. Lower panel: Same pixel by pixel spectra towards the region 2 in the figure \ref{fig:fig3}. Here also highlighted yellow region \textbf{b} is the pixel from where infall velocity is measured.}
 \label{fig:fig5}
\end{figure*}

\begin{figure*}
\includegraphics[width=3.5in,height=3.4in, angle=0]{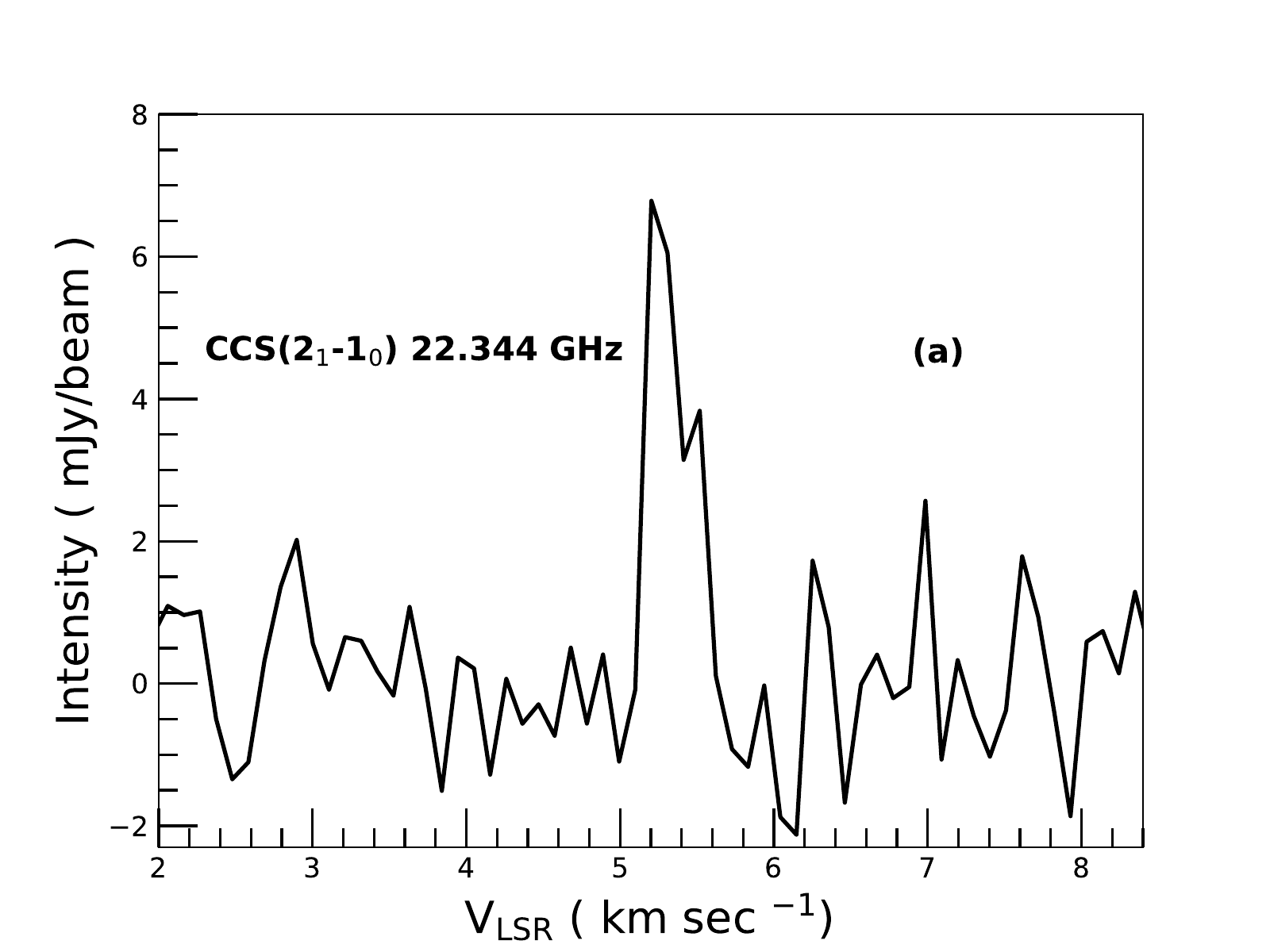}\includegraphics[width=3.5in,height=3.4in,angle=0]{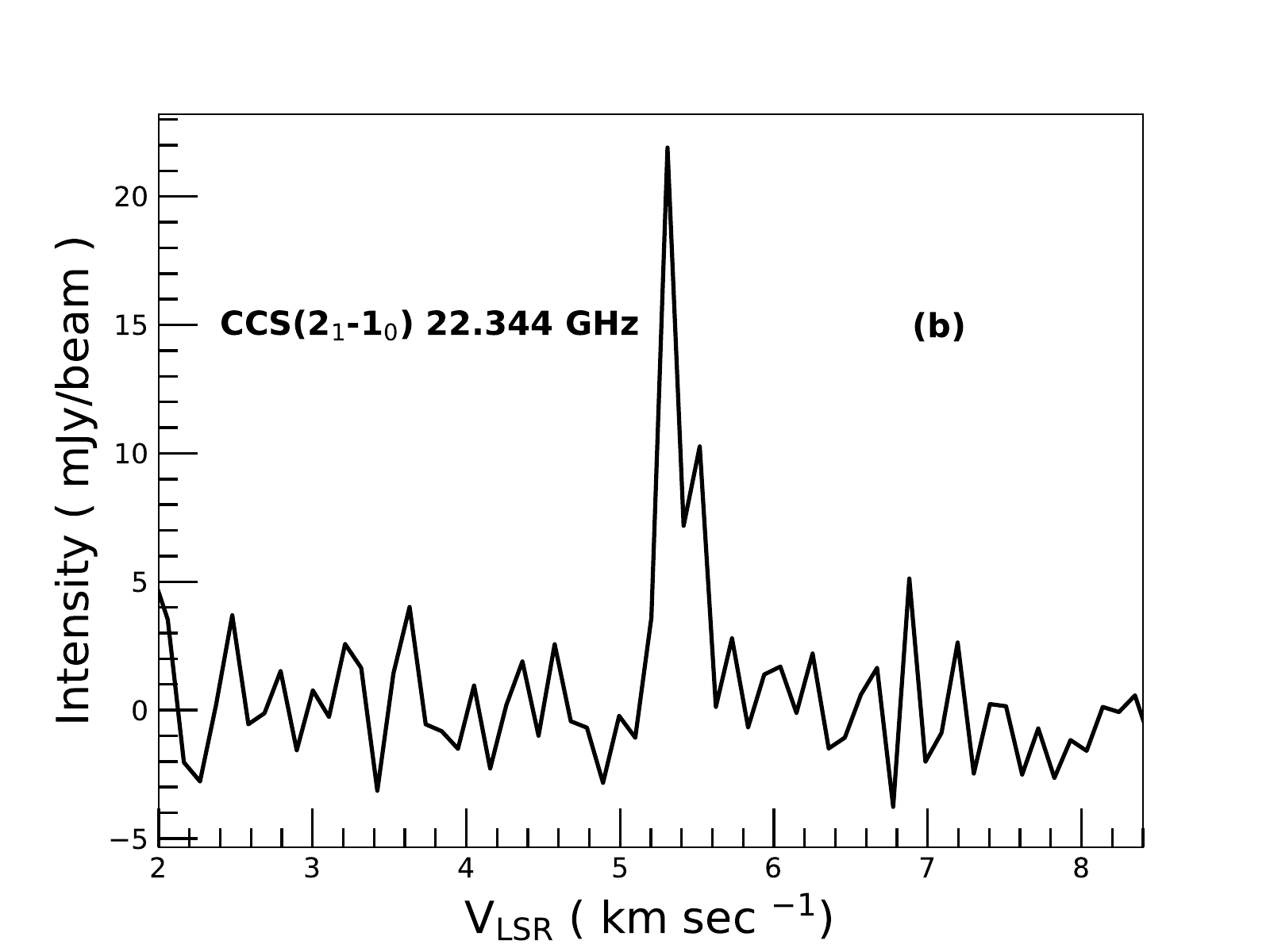}
  \caption{Left Figure: Blue-skewed profile towards the pixel marked by the symbol \textbf{a} in the figure \ref{fig:fig5}.RMS noise of the spectra is $\sim$ 0.66 mJy beam$^{-1}$. Right Figure: Similar type of profile towards the pixel \textbf{b} in figure \ref{fig:fig5}. RMS noise of the spectra is $\sim$ 1.0 mJy beam$^{-1}$.}
 \label{fig:fig6}
\end{figure*}

\begin{figure*}
~~~~~~~~~\textbf{Centre velocity}~~~~~~~~~~~~~~~~~~~~~~~~~~~~~~~~~~~~~~~~\textbf{Velocity dispersion}~~~~~~~~~~~~~~~~~~~~~~~~~~~~~~~~~~~~~~~~~~~~~~~~~~~~~~~~~~~~\\
\includegraphics[width=2.4in,height=2.4in,angle=0]{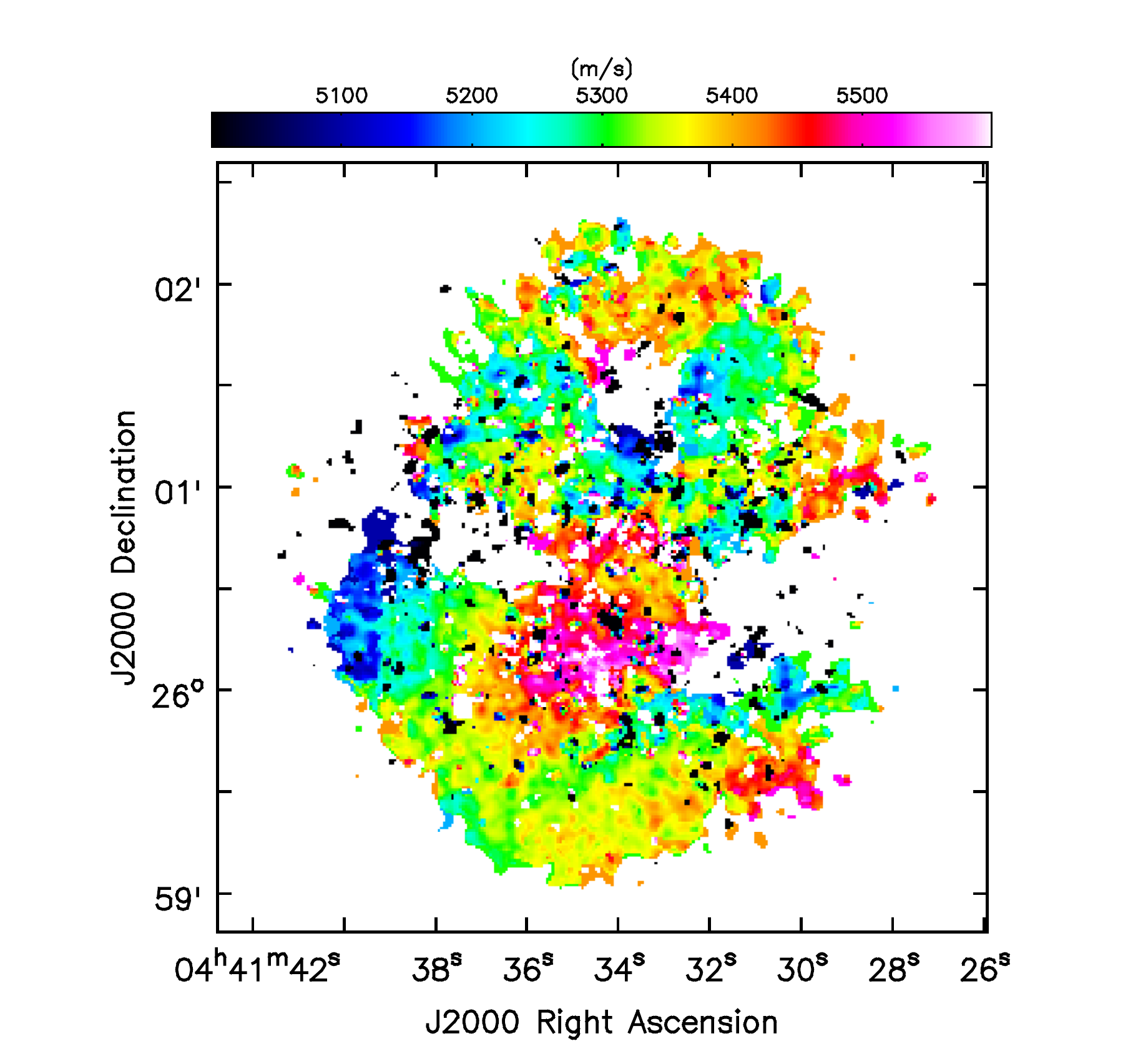}\includegraphics[width=2.4in,height=2.45in,angle=0]{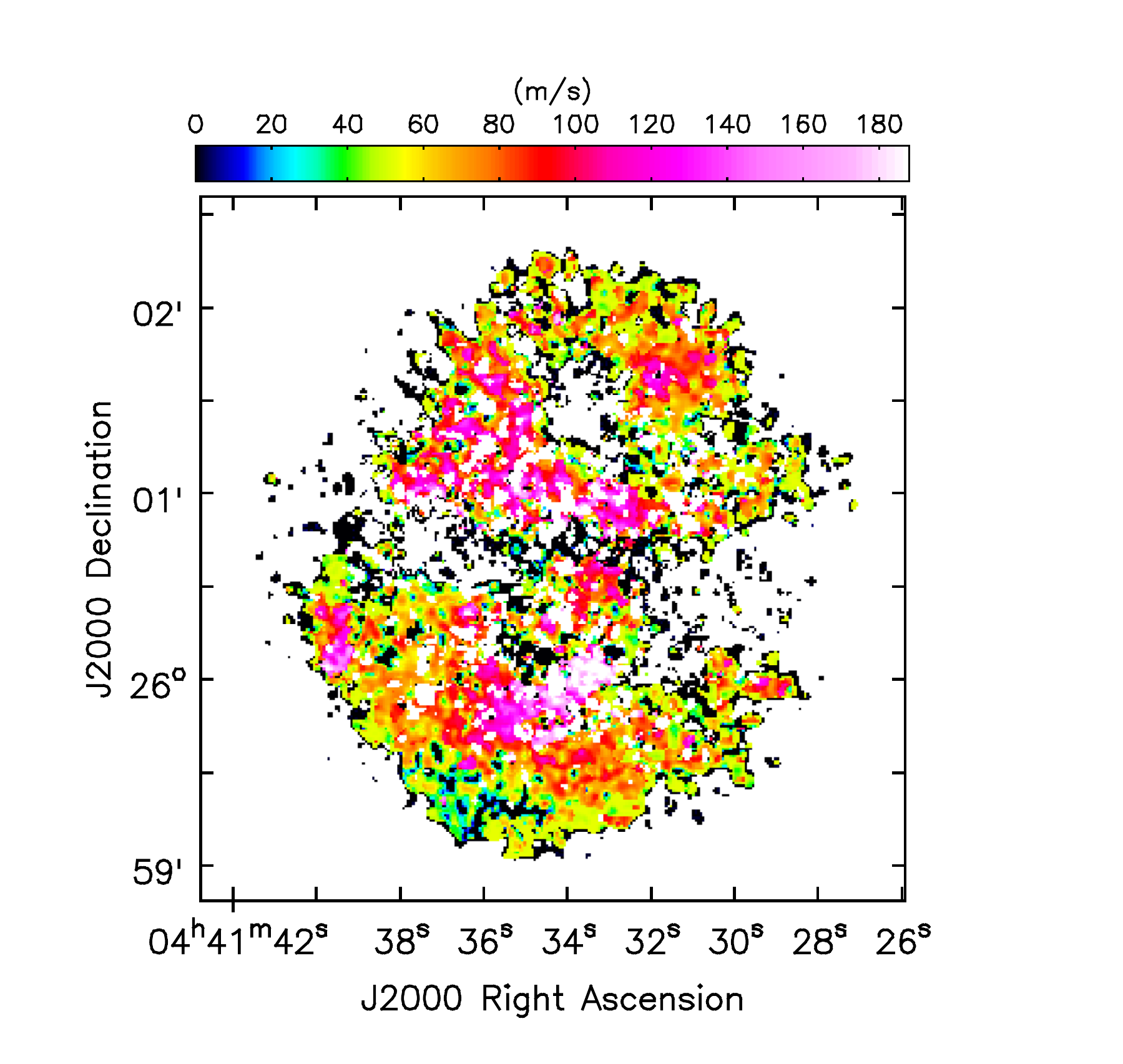}\includegraphics[width=2.4in,height=2.4in,angle=0]{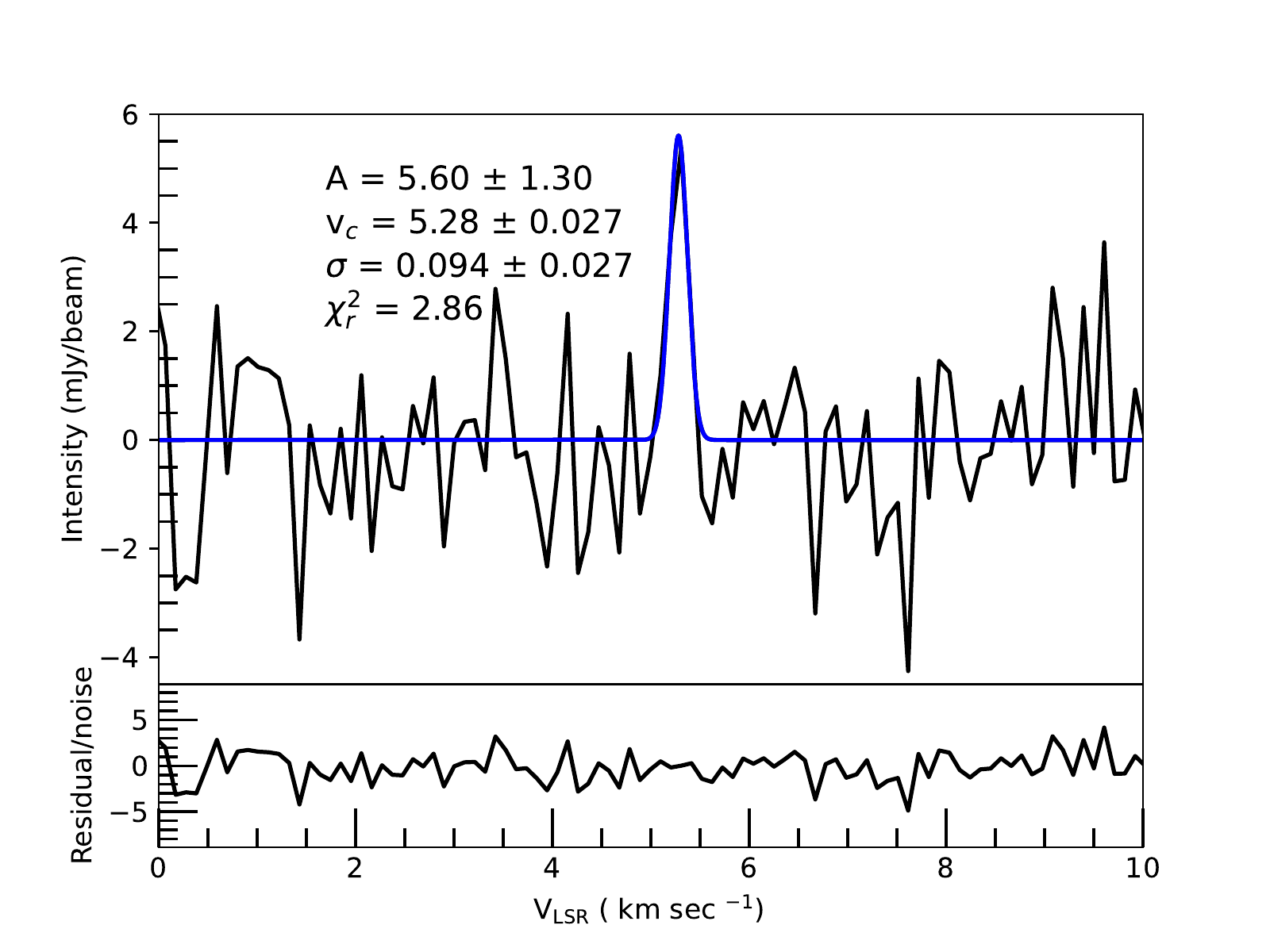}
  \caption{Left Figure: Centre velocity of the CCS (2$_{1}$-1$_{0}$) 22.3 GHz line across the TMC-1C core. Middle Figure: Velocity dispersion of the CCS (2$_{1}$-1$_{0}$) 22.3 GHz line across the core. Right Figure: Spectrum towards the lower right portion of the figure \ref{fig:fig4}, where a single Gaussian type profile is observed ($\alpha = 04^{\rm h}41^{\rm m}31\fs70$,
  $\delta =25^{\circ}59{\arcmin}50\farcs24$). Fitted parameters of the component is mentioned there.}
 \label{fig:fig7}
\end{figure*}

\section{Taurus molecular cloud :}\label{different cores}

Taurus Molecular Cloud is situated in the Taurus and Auriga constellations. The distance of this cloud is 137 pc \citep{2007ApJ...671.1813T}. Taurus molecular cloud is not spherical in shape but rather a sheet-like structure. Using the relation between core velocity dispersion (CVD) and the projected core separation, \citet{2015ApJ...811...71Q} measured the thickness of several well-known molecular clouds, including Taurus. From their analysis, the thickness of the Taurus is $\sim$ 0.7 pc, $\sim$ ten times lesser than the cloud projected length. Also, with the CVD, \citet{2018ApJ...864..116Q} measured the turbulence dissipation energy in Taurus. Taurus is an ideal place for studying the star formation theory because of its nearest distance. Nowadays, this cloud has been extensively studied with different molecules \citep{2021MNRAS.501..347A,2020MNRAS.498.1382W,2020MNRAS.496.4546H,2000ESASP.445..107H}. This molecular cloud has several dense cores; many of those are on the verge of star formation. We have studied three of such cores TMC-1C, TMC-1, and L1544. The phase centers of these sources for these observations are written in table \ref{tablee:table}. Below, we have discussed first the TMC-1C core. After that, at the last part, we have briefly mentioned two other cores, L1544 and TMC-1.\\

\subsection{TMC-1C:}

TMC-1C is a prestellar core in Taurus molecular cloud. This core has been extensively studied with various molecular spectral lines as well as dust continuum \citep{2021A&A...653A..15N,2016A&A...594A.117P,2016A&A...586A.110S, 2011ApJ...739L...4R,2008A&A...487..993K,2007ApJ...657..838S,2007ApJ...671.1839S,1998ApJ...504..207B}.
Several earlier studies argued that it is an evolved starless core. For example, using the Nobeyama 45-meter telescope, \citet{1992ApJ...392..551S} found that the column density ratio of NH$_{3}$ and CCS is $\sim$ 18.38, which clearly indicates that it is an evolved core. Similarly, in different astrochemical models based on the observed N$_{2}$H$^{+}$(J=1-0) abundance or with the observed column density ratios of dense cores' tracers, it was found that the minimum age of TMC-1C is  3 $\times$ 10$^{5}$ yr from  \citet{2007ApJ...671.1839S} and $\sim$ 1 Myr from
 \citet{2021A&A...653A..15N}.

\begin{table}
\caption{Phase centers of TMC-1C, L1544, and TMC-1 cores.}
\centering
\begin{tabular}{|l|c|c|}
\hline
Source & Right Ascension (R.A.) & Declination (Dec.)\\

  & (J2000) & (J2000)\\
  \hline
TMC-1C & $\alpha = 04^{\rm h}41^{\rm m}34\fs31$ &  $\delta =26^{\circ}00{\arcmin}42\farcs59$\\
L1544 & $\alpha = 05^{\rm h}04^{\rm m}15\fs25$ &  $\delta =25^{\circ}11{\arcmin}47\farcs86$\\
TMC-1 & $\alpha = 04^{\rm h}41^{\rm m}42\fs48$ &  $\delta =25^{\circ}41{\arcmin}27\farcs02$\\
\hline
\end{tabular}
\label{tablee:table}
\end{table}

Detailed kinematical and physical studies of this core have been done by \citet{2007ApJ...657..838S, 2007ApJ...671.1839S}, where the authors used the molecular tracers C$^{17}$O (J=1-0), C$^{17}$O (J=2-1), C$^{18}$O (J=2-1), N$_{2}$H$^{+}$(J=1-0), C$^{34}$S (J=2-1), DCO$^{+}$(J=2-1), DCO$^{+}$(J=3-2) and dust continuum maps at three bands 450 $\mu$m, 850 $\mu$m and 1200 $\mu$m. Their studies infer about the temperature of the core, the signature of infall, depletion of carbon chain species towards the dust peak, rotation of the core in some regions, etc. Signature of rotation of this core was prominently observed by  \citet{1998ApJ...504..207B} with NH$_{3}$ (J,K = 1,1) observation. \citet{1998ApJ...504..207B} also showed that inside the coherent core ($\sim$ 0.1 pc), turbulence is subsonic in nature and not correlated with the length scale. They argued that it might be due to the resolution effect of the telescope. Nowadays, this core is being routinely observed with various molecules \citep{2016A&A...594A.117P,2016A&A...586A.110S}.\\


\section{Observation and data analysis}\label{observation and data analysis}

\subsection{CCS 22.3 GHz VLA D configuration data of TMC-1C}
The data for this analysis were taken from the archival data center observed with the Karl G. Jansky Very Large Array (VLA) in February 2013 and March as part of the regular proposal (13A-112) in the K band (18.0 - 26.5 GHz). Bandpass and phase calibrators that were used for this study were 3C147 and J0440+2728, respectively. Spectral resolution for this observation is 0.1 km sec$^{-1}$. Primary beam at this frequency is 135$''$  whereas the synthesized beam is $\sim$ 3.58$''$ $\times$ 3.40$''$. Total on source time for this observation was 13 hr 30 min, and RMS noise that has been achieved $\sim$ 0.8 mJy beam$^{-1}$. All the initial flagging and calibration were done both manually in AIPS (Astronomical Image Processing System; version 31DEC19) and with the scripted CASA (Common Astronomy Software Applications package) pipeline of the National Radio Astronomical Observatory (NRAO)\footnote{The NRAO is operated by Associated Universities Inc., under a collaborative agreement with the US National Science Foundation.}. This was done mainly to check the consistency of the analysis. There was no noticeable difference found between the output of the two methods. We used the multiscaling during the CASA task \textbf{TCLEAN} to pick up all the structures in the core. The scales that we used are 3.$''$2, 7.$''$2, 16.$''$0, 24.$''$0, and 32.$''$0 respectively. Please note that multiscale cleaning produces a single image cube restored with a synthesized beam of the full data.\\

\subsection{1200 $\mu$m dust of TMC-1C:}

Integrated dust emission in TMC-1C at 1200 $\mu$m was taken from the archival data center\footnote{Data available through the Strasbourg astronomical Data Centre (CDS): \url{https://vizier.u-strasbg.fr/viz-bin/VizieR-3?-source=J/A\%2bA/487/993/object}}. It was observed with Max-Planck Millimeter Bolometer (MAMBO)-2 on the IRAM 30-meter telescope. Beam size at this wavelength is 10.80$''$ $\times$ 10.50$''$ and the  RMS noise level is $\sim$ 3 mJy beam$^{-1}$\citep{2008A&A...487..993K}.

\section{Different kinematical and physical properties of TMC-1C:}\label{discussions}

\subsection{CCS distribution and structure:}
Figure \ref{fig:fig1} is the channel maps of CCS (2$_{1}$-1$_{0}$) 22.3 GHz emission in TMC-1C core. These channel maps are without the primary beam correction to show the central part more prominently. From the maps, it is clearly seen that the emission is patchy and widespread. Maybe the large-scale structures are resolved out in interferometric observation. To verify this, we have compared our result with the zero spacing single dish CCS emission taken from the Nobeyama 45-meter telescope \citep{1992ApJ...392..551S}. The integrated spectra of our CCS 22.3 GHz emission is shown in the figure \ref{fig:fig2}. The integrated flux that we have obtained from this study is 5.297 Jy km sec$^{-1}$.
With the conversion factor of 2.8 Jy K$^{-1}$ \citep{2010PASJ...62L..17T}, we obtain the flux of CCS emission recovered by the Nobeyama telescope is 1.932 Jy km sec$^{-1}$. We note that, at this 22.3 GHz frequency, the FWHM of the Nobeyama 45-meter telescope is $\sim$ 75$''$. For a one-to-one comparison, we cut the image cube into a circle whose center is the pointing center and diameter is $\sim$ 75$''$. Then we have obtained the integrated flux is 0.64 Jy km sec$^{-1}$. Thus we only recover  $\sim$ 33\%  of the total integrated flux observed with a single-dish telescope. The rest of the emission is resolved out in the VLA D configuration. These emissions are primarily diffuse large-scale structures ($>$ 66$''$) in favorable conditions for producing CCS.\\
Figure  \ref{fig:fig3} is the overplot of the integrated intensities of CCS (2$_{1}$-1$_{0}$) 22.3 GHz emission and 1200 $\mu$m dust emission. Dust emission is shown in the contour plot, and CCS emission is in the color plot. From the overplot, it shows that there is a depletion of CCS towards the dust peak. The CCS distribution that surrounds the cores seems to be consistent with the previous studies shown by \cite{2014ApJ...789...83T,2001ApJ...552..639A}, and \cite{2000ApJS..128..271L}. Depletion of other carbon-bearing species like C$^{17}$O(J=1-0), C$^{18}$O(J=2-1), CS(J=3-2), etc., towards the dust peak is also observed by the previous studies in TMC-1C \citep{2007ApJ...671.1839S}. \citet{2007ApJ...671.1839S} also showed that the depletion factor of the carbon-bearing species like C$^{17}$O(J=1-0), C$^{18}$O(J=2-1) has a nice correlation with H$_{2}$ column density and anticorrelation with dust temperature. Two red boxes in this figure are the regions where pixel-by-pixel spectra are shown in the next section for measuring the infall velocity. The black square box is the region where we have tentatively detected the magnetic field (Koley et al. 2022 in prep.).

\subsection{Spectral properties across the core:}
Spectral properties across the core are very complex. In figure \ref{fig:fig4}, we have shown the spectra across TMC-1C after binning it into 15 $\times$ 15 regions. For most of the region, the self-absorbed profile is observed. It indicates that the emission is optically thick. In some places, mainly far from the dust peak, the profile looks like a single Gaussian, whereas in few regions, a separate weak component is observed. In some places, it seems to be a classic blue-skewed profile (specifically towards the red boxes in figure \ref{fig:fig4}). We have modeled such infall profile towards the region \textbf{1} and \textbf{2} mentioned in figure \ref{fig:fig3}. We note that, towards the dust peak, CCS emission is less. Thus, maybe due to the sensitivity issue, we have not prominently observed the infall signature towards this region through CCS radical. However, \citet{2007ApJ...671.1839S} observed the infall velocity towards the dust peak through N$_{2}$H$^{+}$(J=1-0) emission. This type of profile can be modeled with two layers of a collapsing region
\citep{2012ApJ...759L..37C,2005ApJ...620..800D,2004ApJS..153..523L,1999ARA&A..37..311E,1996ApJ...465L.133M} and from the model, it is possible to measure the infall velocity v$_{in}$ of the collapsing system. We discuss about this in detail in the next section.\\

\subsection{Infall velocity:}\label{infall velocity}
Molecular cloud evolves from translucent, relatively low-density environments and forms the dense cold-core.
Suppose magnetic field and turbulence are high in the initial stage of cloud evolution. In that case, these will reduce during the evolutionary stage. Magnetic flux is reduced through the ambipolar diffusion process \citep{1956MNRAS.116..503M}. In contrast, turbulence is reduced through dissipation at smaller scales \citep{1999ApJ...526..279P,1998ApJ...504..207B}. When these two opposing forces have gone, the dense material collapse on its own. Suppose some region of the core is in a stage of collapse. In that case, the signature of this is observed in the molecular spectra. This type of spectra can be modeled as two layers of the collapsing region.
Signature of the collapse was found earlier in TMC-1C  studied by  \citet{2007ApJ...671.1839S} from the tracer N$_{2}$H$^{+}$(J=1-0). Their analysis found the signature of infall towards the dust peak and the northwest regions of the dust peak. Near the dust peak, infall speed is $\sim$ 0.15 km sec$^{-1}$, and towards the northwest region, it is $\sim$ 0.05 km sec$^{-1}$. Infall velocity can be measured from the observed parameters of the spectra by the formulae \citep{1996ApJ...465L.133M}:

\begin{eqnarray}\label{eq:equation2}
V_{in} \approx  \displaystyle\frac{\sigma_{thin}^{2}}{V_{red}-V_{blue}} ln \left(\frac{1 + e^{\frac{I_{BD}}{I_{D}}}}{1 + e^{\frac{I_{RD}}{I_{D}}}} \right)
\end{eqnarray}

Here, V$_{in}$ is the infall velocity, V$_{red}$ and V$_{blue}$ are the center velocities of the red and the blue peaks. $\sigma_{thin}$ is the velocity dispersion of the optically thin line, I$_{D}$ is the intensity of the dip in between the two peaks, I$_{BD}$ is the difference of intensity between the blue peak and the dip, whereas I$_{RD}$ is the difference of intensity between the red peak and the dip. \\
We have modeled the spectra towards the pixels \textbf{a} and \textbf{b} in figure \ref{fig:fig5} (individual spectra are in figure \ref{fig:fig6}) with the two layers model of a collapsing region. As we did not observe any co-existing optically thin line, we have taken the $\sigma_{thin}$ = 0.10 km sec$^{-1}$ after fitting the spectrum, which is far from the dust peak where the self-absorption effect is less (detailed discussion in the next section).
Thus, from the equation \ref{eq:equation2}, we have obtained the infall velocities 0.035 and 0.06 km sec$^{-1}$ towards the pixels \textbf{a} and \textbf{b}. The variation of the velocity is due to the projection effect. Because, from this spectral line modeling, we only observe the line-of-sight component of the total infall velocity.\\ 
Now, we compare this infall velocity with the free-fall velocity of the collapse. 
Free-fall velocity is obtained from the formulae: $\displaystyle v_{ff} =\sqrt\frac{2GM}{R}$. Here, M is the mass of the core, R is the radius of the core and G is the gravitational constant. The mass of the TMC-1C core is $\sim$ $\textup{M}_\odot$ (\cite{2021A&A...653A..15N}, and references therein) and size is $\sim$ 0.1 pc \citep{1998ApJ...504..207B}. Thus, the free-fall velocity is $\sim$ 0.28 km sec$^{-1}$.
We note that the infall velocity obtained from the spectral line modeling is the line-of-sight component of the total infall velocity. Thus, after taking into the projection effect, it will be more or less similar to the free-fall velocity.

\subsection{Non-thermal velocity dispersion:}
Thermal velocity dispersion of CCS is:
\begin{eqnarray}\label{eq:equation1}
\displaystyle \sigma_{th}=\sqrt\frac{k_{B}T_{k}}{m_{ccs}}
\end{eqnarray}

Here, $\sigma_{th}$ is the thermal velocity dispersion, k$_{B}$ is Boltzmann constant = 1.38 $\times$ 10$^{-23}$ J/K, T$_{k}$ is the kinetic temperature and m$_{ccs}$ is the mass of the CCS radical.
From the dust maps in three bands (450$\mu$m, 850$\mu$m, and 1200$\mu$m), \citet{2007ApJ...657..838S} studied the temperature of the core and found that the temperature of the dense nucleus of the core is as low as 6K. As CCS radical traces slightly outsider region relative to the dust, in our analysis, we assume the kinetic temperature T$_{k}$ is 10K. Then from equation \ref{eq:equation1}, $\sigma_{th}$ is 0.038 km sec$^{-1}$. Centre velocity distribution and the velocity dispersion of the spectra are shown in the left and middle panels of figure \ref{fig:fig7}. From our integrated velocity dispersion map, we see the average total velocity dispersion, $\sigma_{tot}$ is 0.1 km sec$^{-1}$. 

However, as much of the spectra are affected by the self-absorption effect, the average velocity dispersion taken from this moment map is incorrect. Thus, to obtain an accurate velocity dispersion, we have taken the spectrum towards the right lower panel of figure \ref{fig:fig4}, where a single Gaussian type profile is observed and assumed to be less affected by the self-absorption effect. From the overplot in figure \ref{fig:fig2}, it is clear that this region is far from the central core. Thus, the velocity dispersion taken from such a spectrum can be considered an accurate velocity dispersion of the spectrum. We have taken one such spectrum and fitted it with a single Gaussian. This spectrum is shown in the right panel of figure \ref{fig:fig7}. From the fitted component, we have observed the total velocity dispersion of the spectrum is $\sim$ 0.1 km sec$^{-1}$. As in our study, the spectral resolution was $\sim$ 0.1 km sec$^{-1}$, we can consider it as an upper limit of the total velocity dispersion. We will continue our further analysis with this upper limit. Now, the non-thermal velocity dispersion $\sigma_{nth}$ is 0.093 km sec$^{-1}$. This is $\sim$ 2.4 times higher than the thermal velocity dispersion. In a molecular cloud major component is H$_{2}$. For, H$_{2}$ molecule, at 10K thermal velocity dispersion is $\sim$ 0.2 km sec$^{-1}$. Turbulence velocity dispersion ($\sigma_{nth}$) obtained from CCS is $\sim$ two times lesser than thermal velocity dispersion of H$_{2}$. Turbulence is thus subsonic in nature inside the core. There are arguments in the literature that turbulence is dissipated near the ambipolar diffusion scale. After that, a coherent core of $\sim$ 0.1 pc is formed, where turbulence is subsonic and has no dependence on the length scale \citep{1998ApJ...504..207B}.
Similar kinds of studies have also been done by \citet{2007ApJ...671.1839S} with different molecules (both carbon and nitrogen-bearing species) towards the dust continuum peak. Their analysis showed that, after taking the gas temperature of 10K, the ratio between non-thermal and thermal velocity dispersion is $\sim$2  for NH$_{2}^{+}$(J=1-0) and $\sim$ 3-4 for C$^{17}$O(J=1-0) and C$^{18}$O(J=2-1). They attributed that this occurs because later ones trace the outer relatively low-density part where the former trace the innermost dense region.\\

\subsection{Virial analysis:}
It is useful to calculate, whether thermal and non-thermal motions are able to support the collapse or not. For that we calculate the virial parameter $\alpha$. It is defined by the formulae: $\displaystyle\frac{5{\sigma_{v}}^{2}R}{GM}$ \citep{2013ApJ...779..185K}. Here, $\sigma_{v}$ is the total velocity dispersion due to the thermal and non-thermal motions of the bulk gas. R is radius of the core, G is the gravitational constant and M is the mass of the core. The total velocity dispersion $\sigma_{v}$ for H$_{2}$ molecule at 10K temperature is $\displaystyle\sqrt{0.2{^2} + 0.093^{2}}$ = 0.22 km sec$^{-1}$. For TMC-1C, the typical size is $\sim$0.1 pc \citep{1998ApJ...504..207B}. After converting the mass in the form of column density, we finally obtain $\alpha$ $\approx$ 10$^{22}$ $\times$ ($\displaystyle{\frac{N(H_{2})}{cm^{-2}})}^{-1}$. From the Nobeyama 45 meter telescope, column density of CCS in TMC-1C is 3.1 $\times$ 10$^{13}$ cm $^{-2}$. \citet{2007ApJ...671.1813T} predicted the fractional abundance of $\displaystyle\frac{n(CCS)}{n(H_{2})}$ is (2-6) $\times$ 10$^{-10}$.  Thus, N(H$_{2}$) in the range 5-15 $\times$ 10$^{22}$ cm$^{-2}$. On the other hand, \cite{2008A&A...492..703C} reported the peak column density towards the dust peak is 8.5 $\times$ 10$^{22}$ cm$^{-2}$. From the overplot in figure \ref{fig:fig3}, if we assume that the average CCS is located where the dust emission is half of the peak value, the  column density of  N(H$_{2}$) is 4.25 $\times$ 10$^{22}$ cm$^{-2}$. Moreover, in any case, the value of $\alpha$ is $<$ 1. Thus, thermal and non-thermal motions cannot prevent the collapse. This is consistent with the infall motion found in some portions of the core through CCS radical. Further consideration of the magnetic field in the virial analysis may change the scenario.

\begin{figure*}
~~~~~~~~~~~~~~~~~~~~~~~~~~~~~~~~~~~~~~~~~~~~~~~~~~~~~~~~~~~~~~~~~~~~~~~~~~~~~~~~~~~~~~~~~~~~~~~~~~~~~~~~~~~~~~~~~~~~~~~~~~~~~~~~~~~~~~~~~ Jy/beam. m/sec~~~~~~~~~~~~~~~~~~~~~~~\\
\includegraphics[width=3.6in,height=3.4in,angle=0]{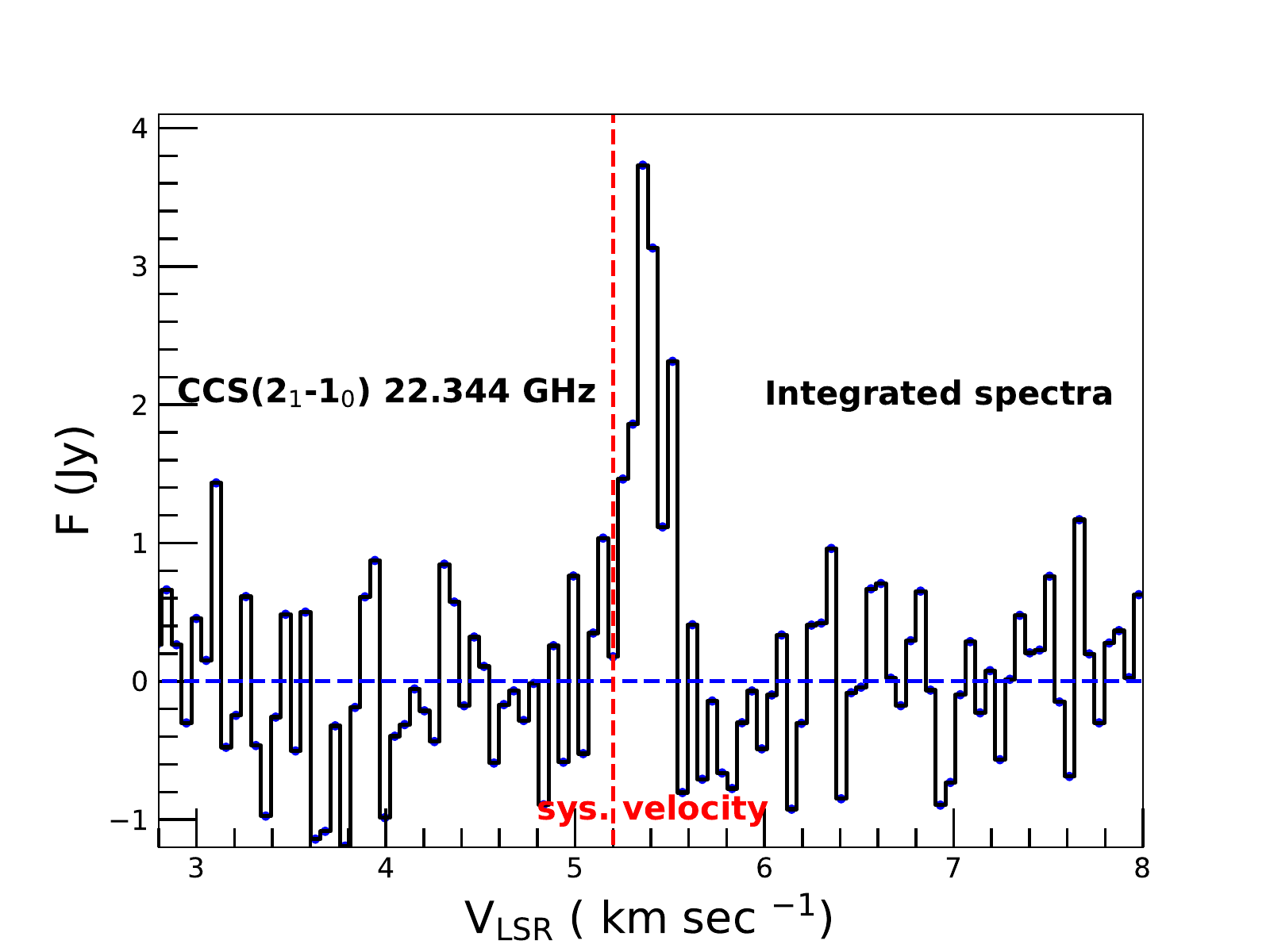}\includegraphics[width=3.3in,height=3.2in,angle=0]{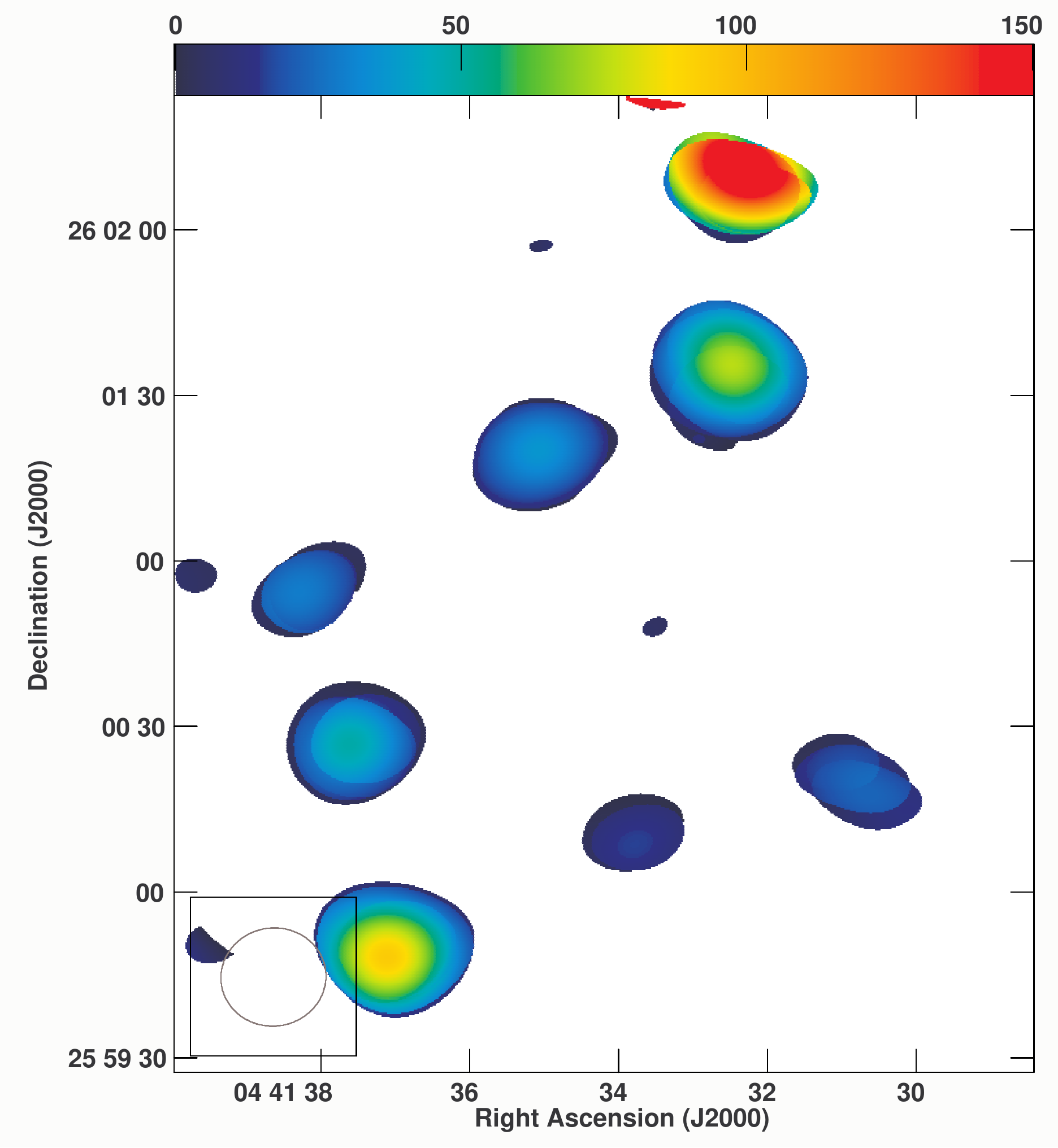}\\
  \caption{(a) Integrated spectra of CCS(2$_{1}$-1$_{0}$)  22.3 GHz line towards the TMC-1C core. RMS noise of the spectra is $\sim$ 0.34 Jy. (b) Integrated intensity map of CCS(2$_{1}$-1$_{0}$)  22.3 GHz line emission towards the TMC-1C core at a angular resolution of 19.05$''$ $\times$ 17.77$''$.}
 \label{fig:fig8}
\end{figure*} 

\begin{figure*}
~~~~~~~~~~~~~~~~~~~~~~~~~~~~~~~~~~~~~~~~~~~~~~~~~~~~~~~~~~~~~~~~~~~~~~~~~~~~~~~~~~~~~~~~~~~~~~~~~~~~~~~~~~~~~~~~~~~~~~~~~~~~~~~~~~~~~~~~~ Jy/beam. m/sec~~~~~~~~~~~~~~~~~~~~~~~\\
\includegraphics[width=3.6in,height=3.4in,angle=0]{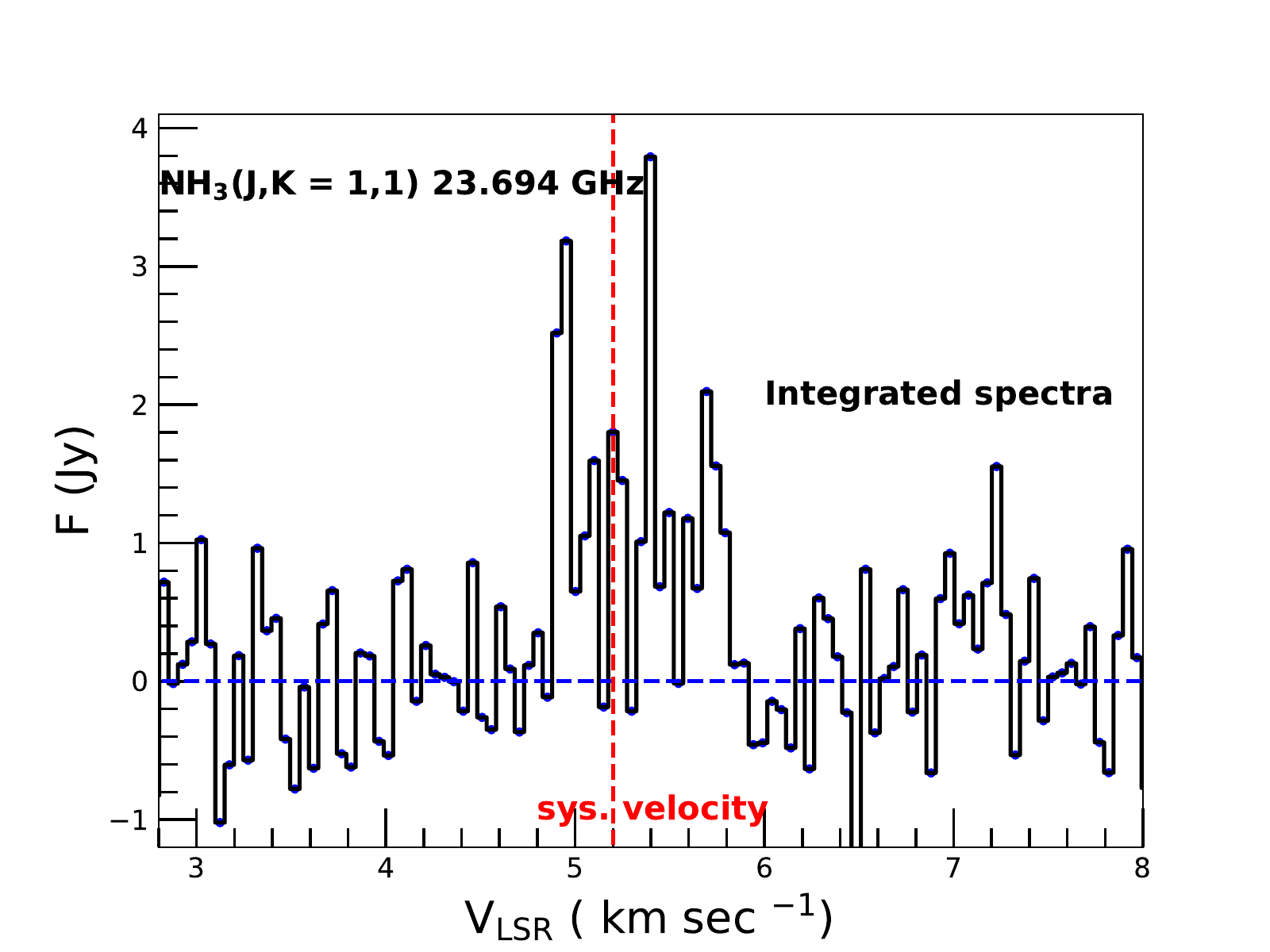}\includegraphics[width=3.3in,height=3.2in,angle=0]{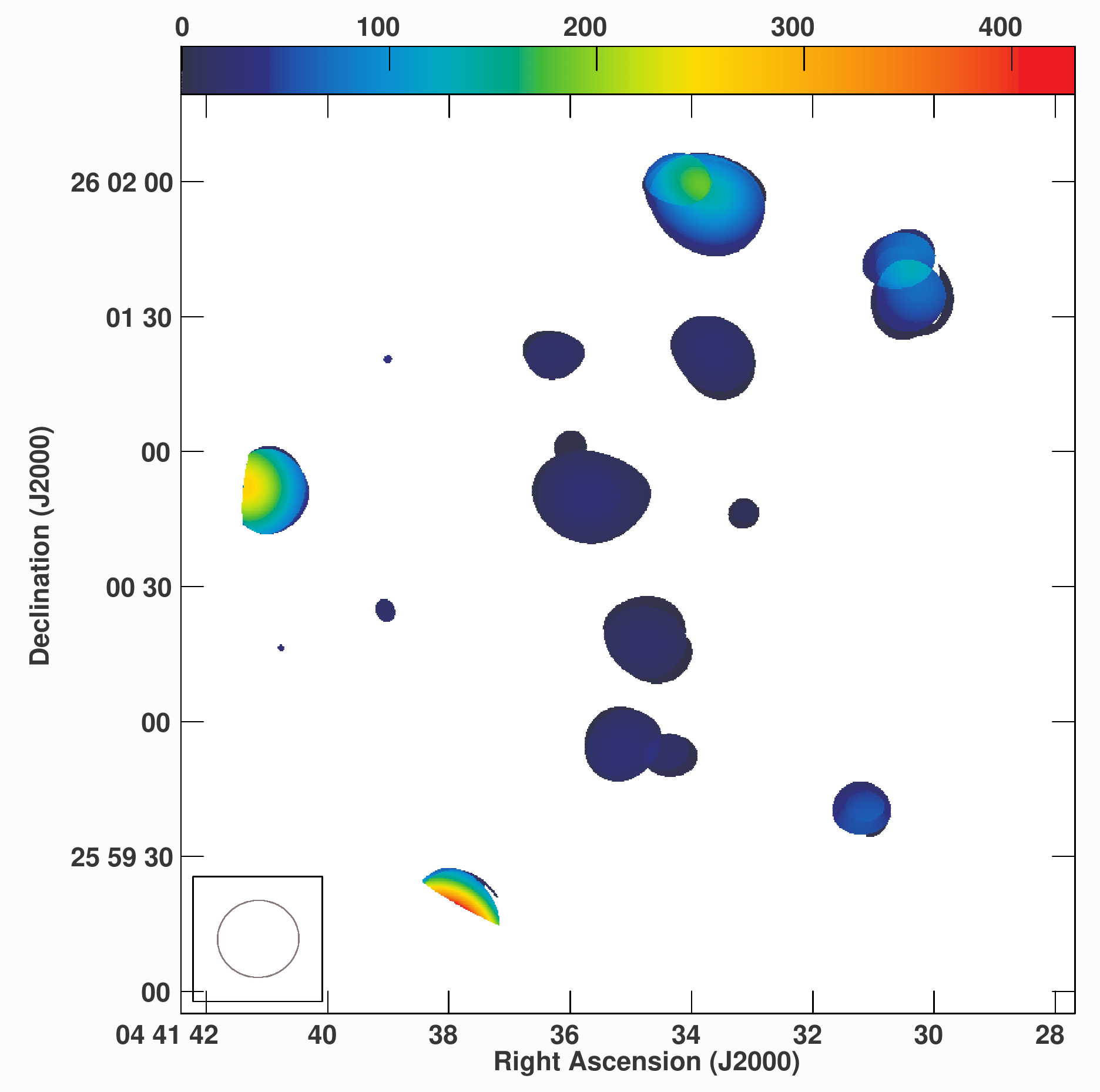}\\

  \caption{Integrated spectra of NH$_{3}$(J,K = 1,1) 23.69 GHz line towards the TMC-1C core. RMS noise of the spectra is $\sim$ 0.34 Jy. (b) Integrated intensity map of  NH$_{3}$(J,K = 1,1) 23.69 GHz line emission towards the TMC-1C core at a angular resolution of 18.10$''$ $\times$  17.11$''$.}
 \label{fig:fig9}
\end{figure*}

\begin{figure*}
~~~~~~~~~~~~~~~~~~~~~~~~~~~~~~~~~~~~~~~~~~~~~~~~~~~~~~~~~~~~~~~~~~~~~~~~~~~~~~~~~~~~~~~~~~~~~~~~~~~~~~~~~~~~~~~~~~~~~~~~~~~~~~~~~~~~~~~~~ Jy/beam. m/sec~~~~~~~~~~~~~~~~~~~\\
\includegraphics[width=3.5in,height=3.4in,angle=0]{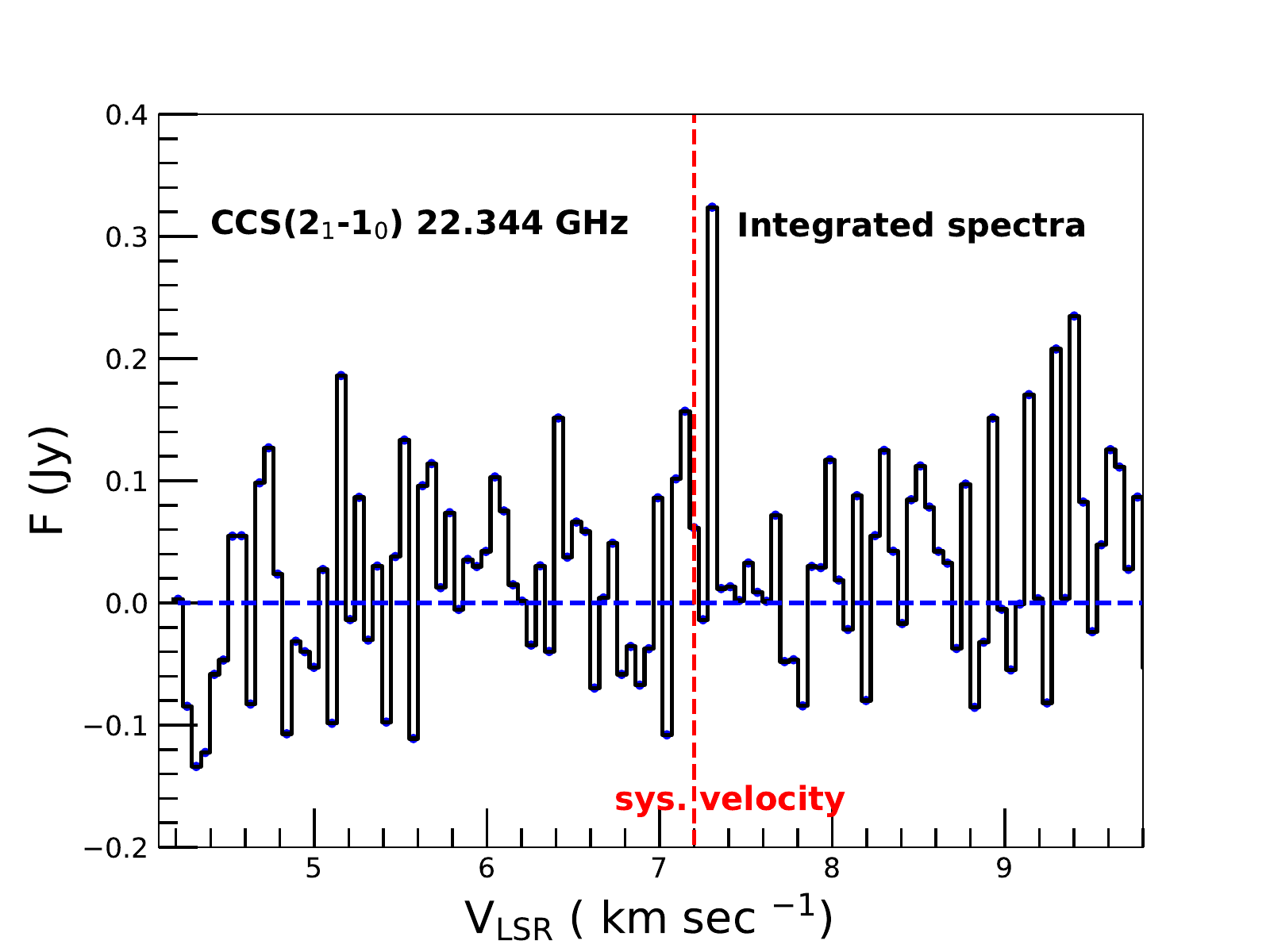}\includegraphics[width=3.5in,height=3.2in,angle=0]{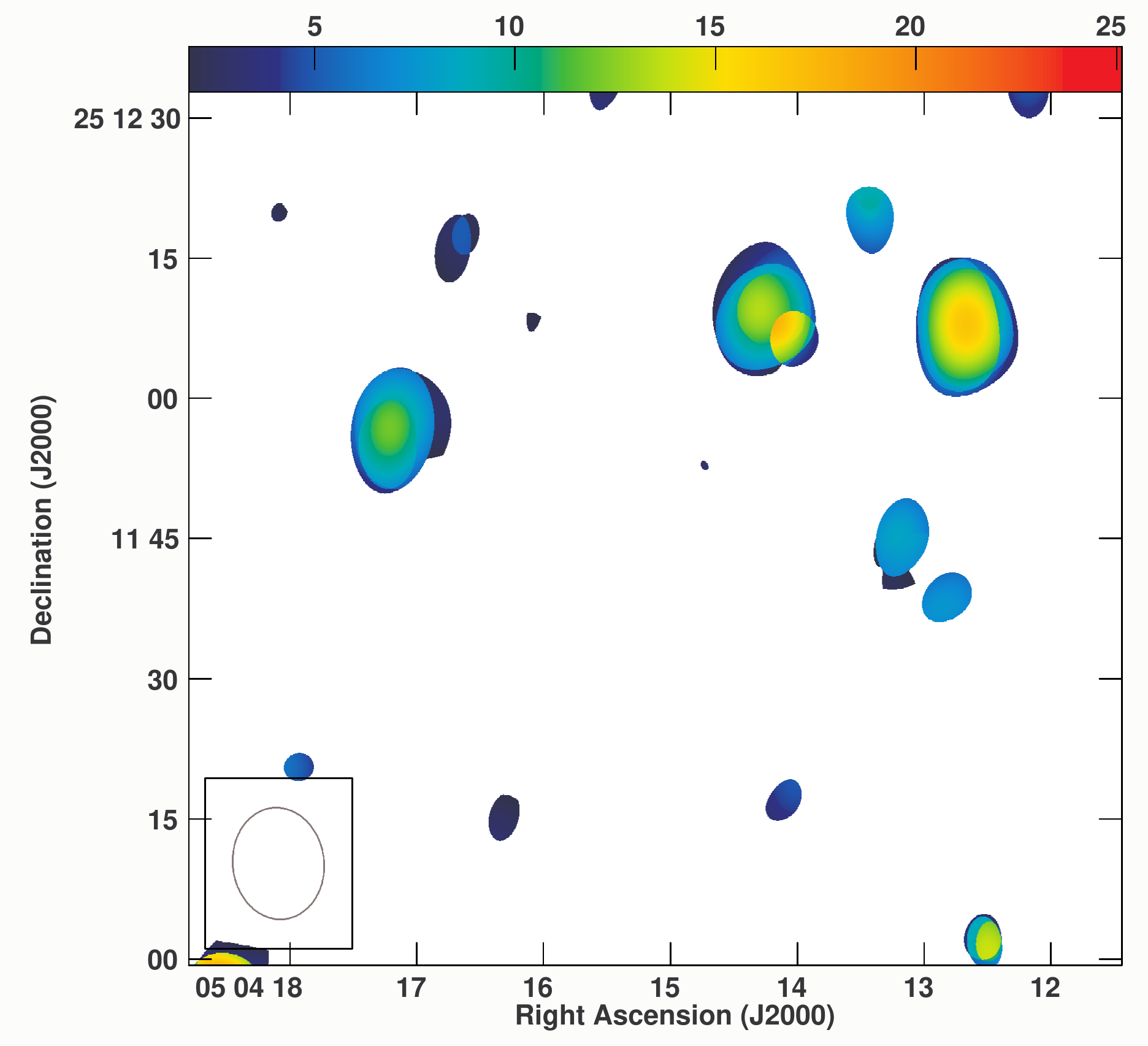}
  \caption{(a) Integrated spectra of CCS (2$_{1}$-1$_{0}$) 22.3 GHz line towards the L1544 core. RMS noise of the spectra is $\sim$ 0.04 Jy.(b) Integrated intensity map of CCS (2$_{1}$-1$_{0}$) 22.3 GHz line emission towards the L1544 core at a angular resolution of 11.96$''$ $\times$ 9.77$''$.}
 \label{fig:fig10}
\end{figure*}

\begin{figure*}
~~~~~~~~~~~~~~~~~~~~~~~~~~~~~~~~~~~~~~~~~~~~~~~~~~~~~~~~~~~~~~~~~~~~~~~~~~~~~~~~~~~~~~~~~~~~~~~~~~~~~~~~~~~~~~~~~~~~~~~~~~~~~~~~~~~~~~~~~ Jy/beam. m/sec~~~~~~~~~~~~~~~~~~~~~~~\\
\includegraphics[width=3.5in,height=3.4in,angle=0]{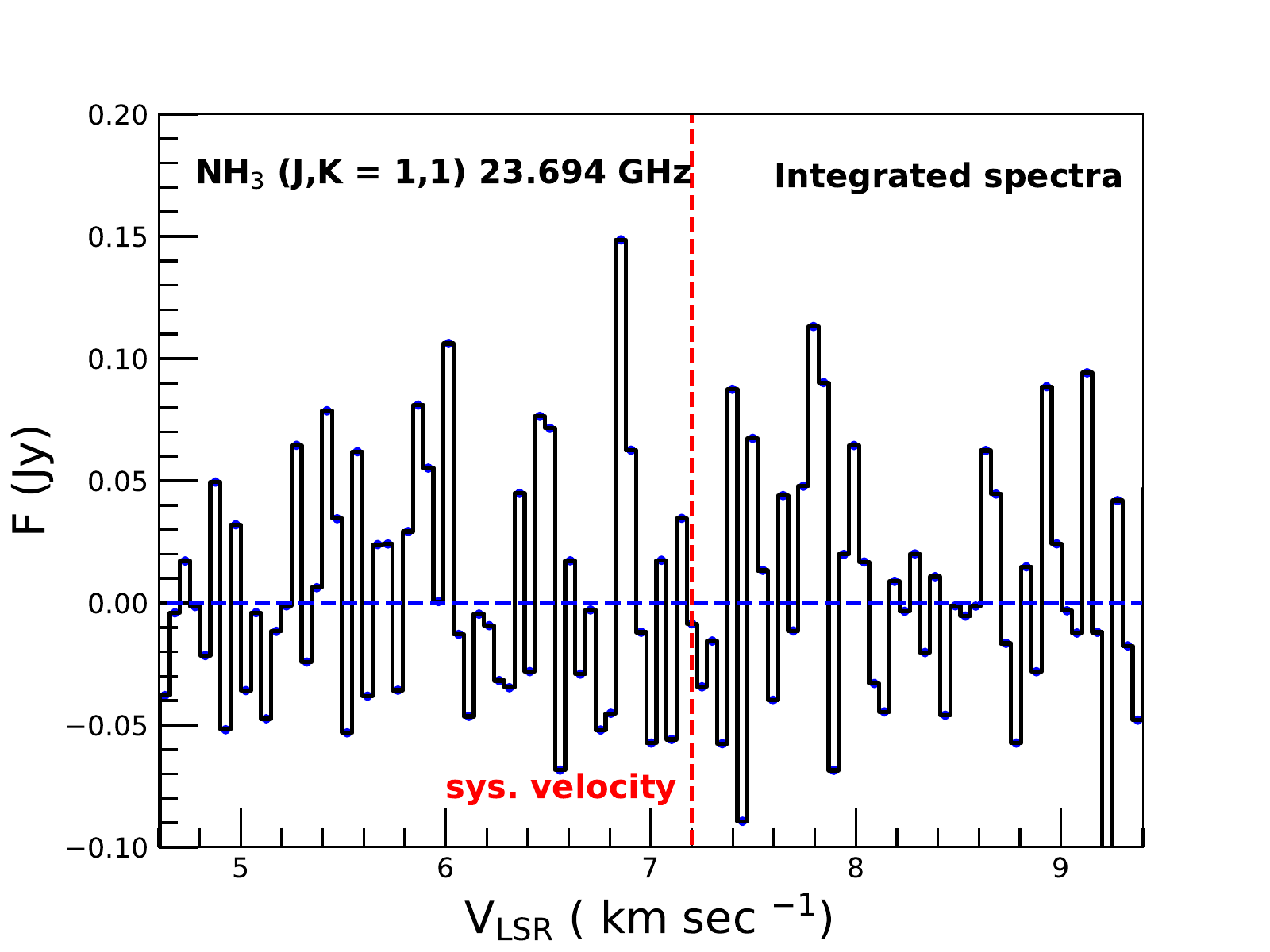}\includegraphics[width=3.5in,height=3.2in,angle=0]{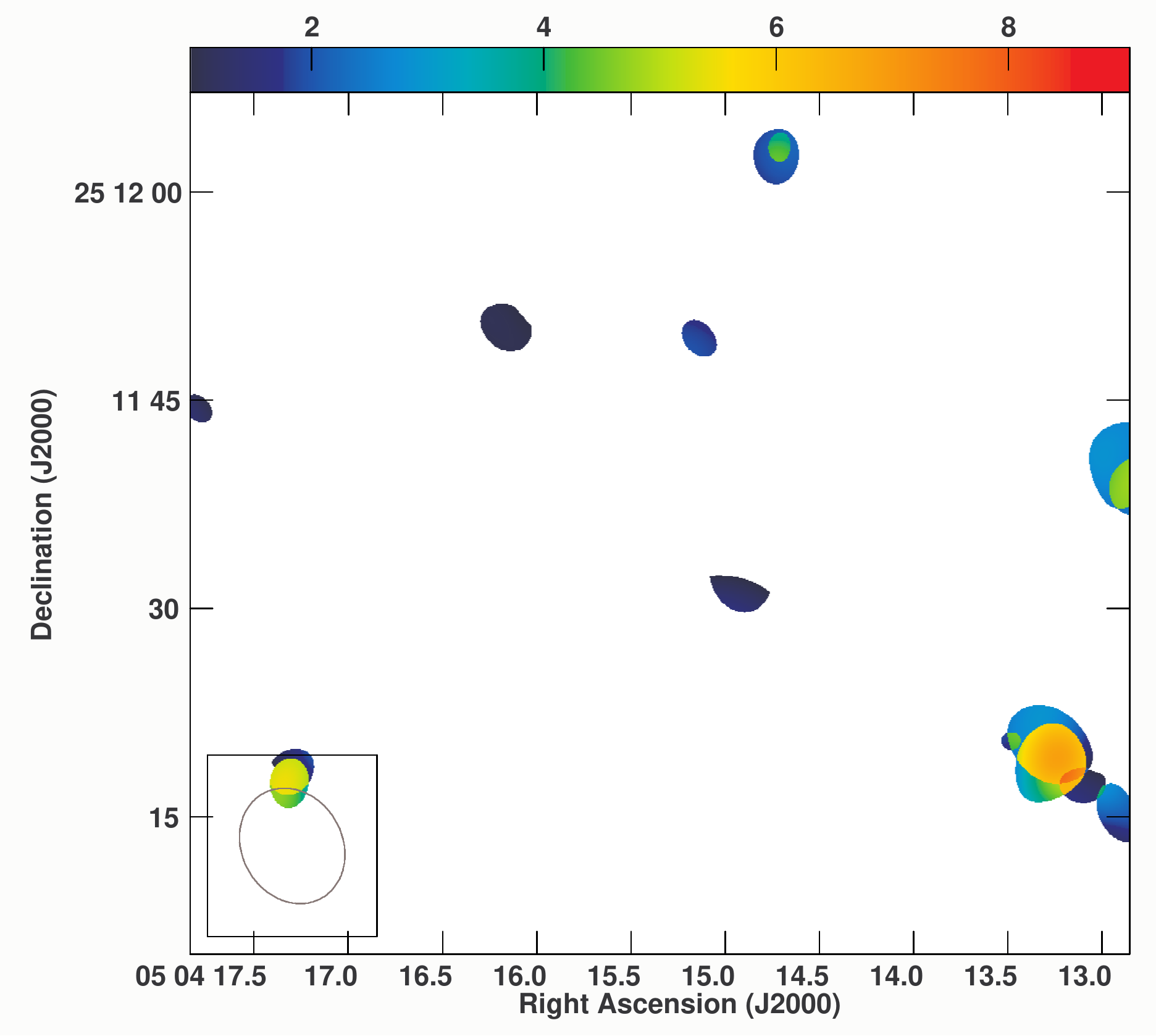}\\
  \caption{Integrated spectra of NH$_{3}$(J,K = 1,1) 23.69 GHz line towards the L1544 core. RMS noise of the spectra is $\sim$ 0.03 Jy. (b) Integrated intensity map of  NH$_{3}$(J,K = 1,1) 23.69 GHz line emission towards the L1544 core at a angular resolution of 8.63$''$ $\times$ 7.23$''$.}
 \label{fig:fig11}
\end{figure*}

\begin{figure*}
~~~~~~~~~~~~~~~~~~~~~~~~~~~~~~~~~~~~~~~~~~~~~~~~~~~~~~~~~~~~~~~~~~~~~~~~~~~~~~~~~~~~~~~~~~~~~~~~~~~~~~~~~~~~~~~~~~~~~~~~~~~~~~~~~~~~~~~~~ Jy/beam. m/sec~~~~~~~~~~~~~~~~~~~~~~~\\
\includegraphics[width=3.5in,height=3.4in,angle=0]{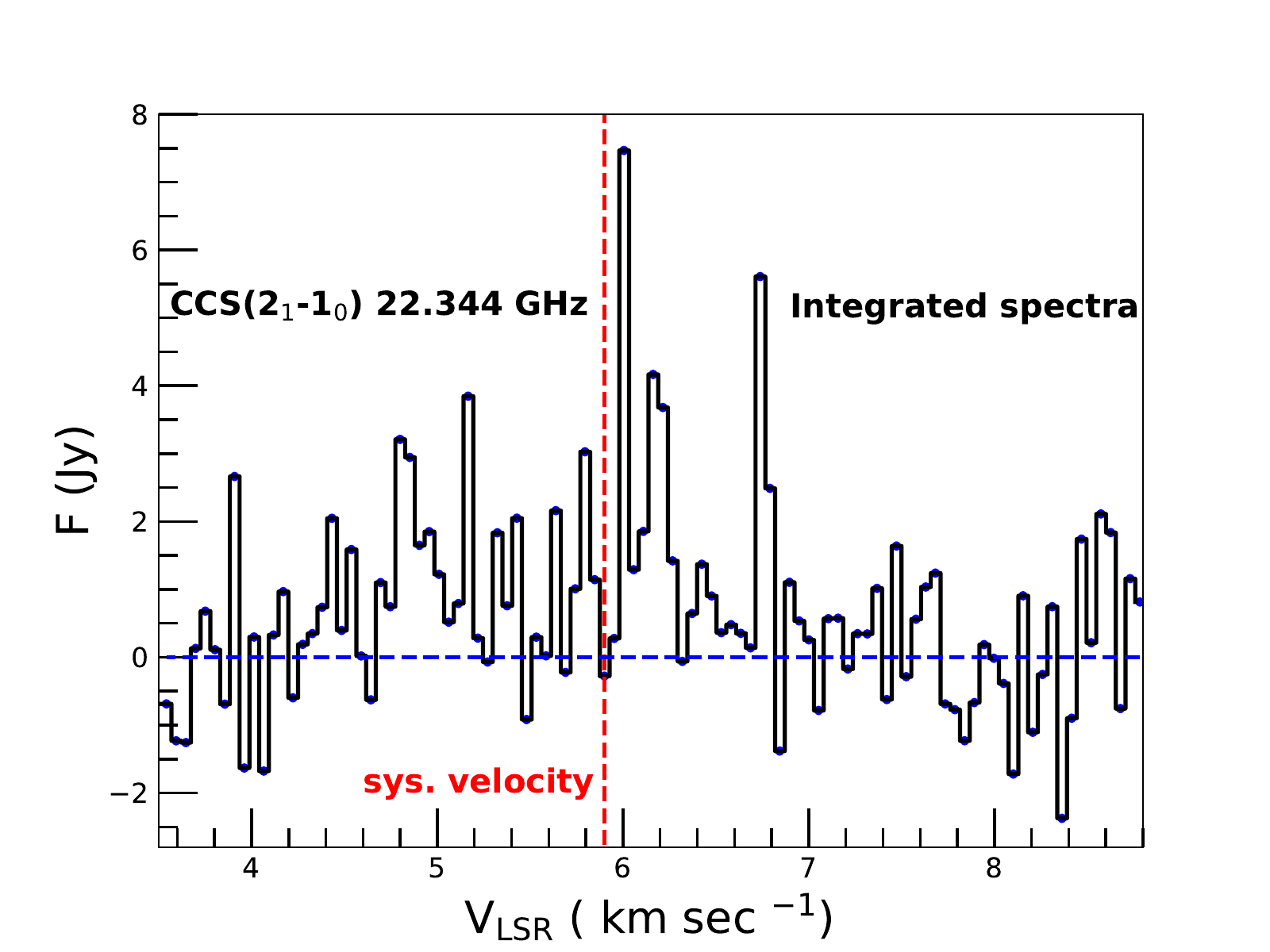}\includegraphics[width=3.5in,height=3.2in,angle=0]{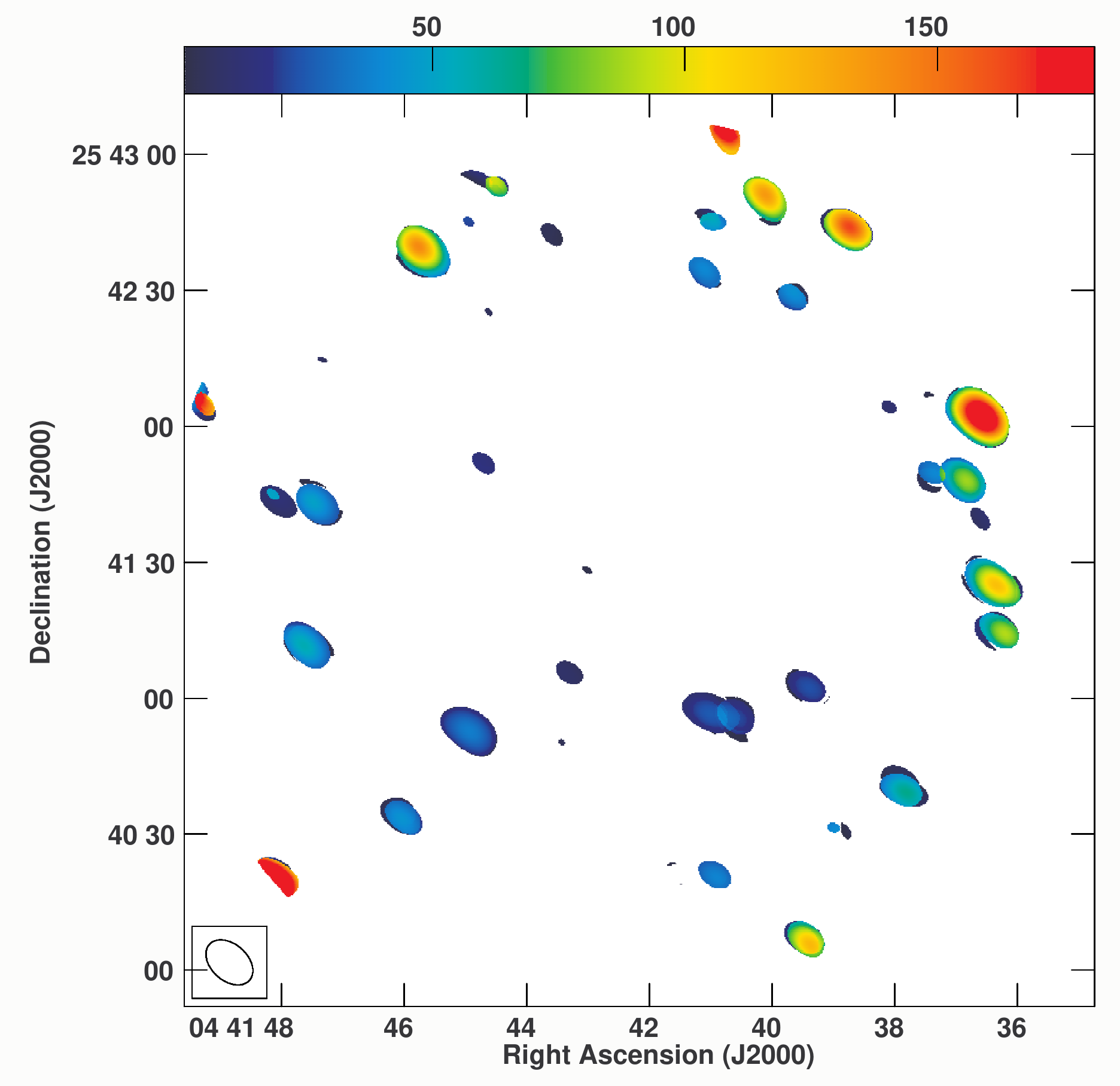}\\
  \caption{(a) Integrated spectra of CCS (2$_{1}$-1$_{0}$) 22.3 GHz line towards the TMC-1 core. RMS noise of the spectra is $\sim$ 0.77 Jy. (b) Integrated intensity map of CCS (2$_{1}$-1$_{0}$) 22.3 GHz line emission towards the TMC-1 core at a angular resolution of 11.87$''$ $\times$ 8.04$''$.}
 \label{fig:fig12}
\end{figure*}

\begin{figure*}
~~~~~~~~~~~~~~~~~~~~~~~~~~~~~~~~~~~~~~~~~~~~~~~~~~~~~~~~~~~~~~~~~~~~~~~~~~~~~~~~~~~~~~~~~~~~~~~~~~~~~~~~~~~~~~~~~~~~~~~~~~~~~~~~~~~~~~~~~ Jy/beam. m/sec~~~~~~~~~~~~~~~~~~~~~~~\\
\includegraphics[width=3.5in,height=3.4in,angle=0]{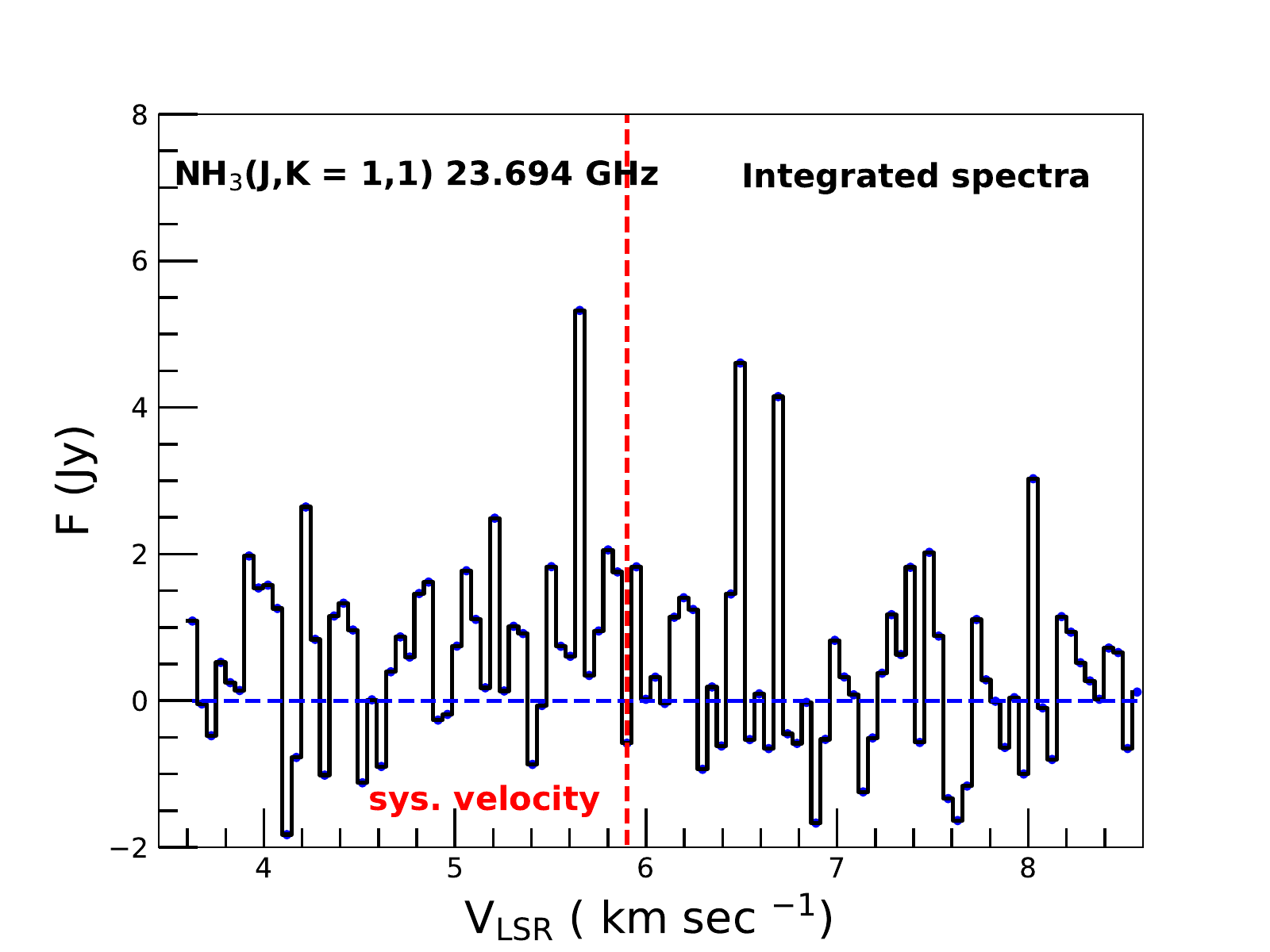}\includegraphics[width=3.5in,height=3.2in,angle=0]{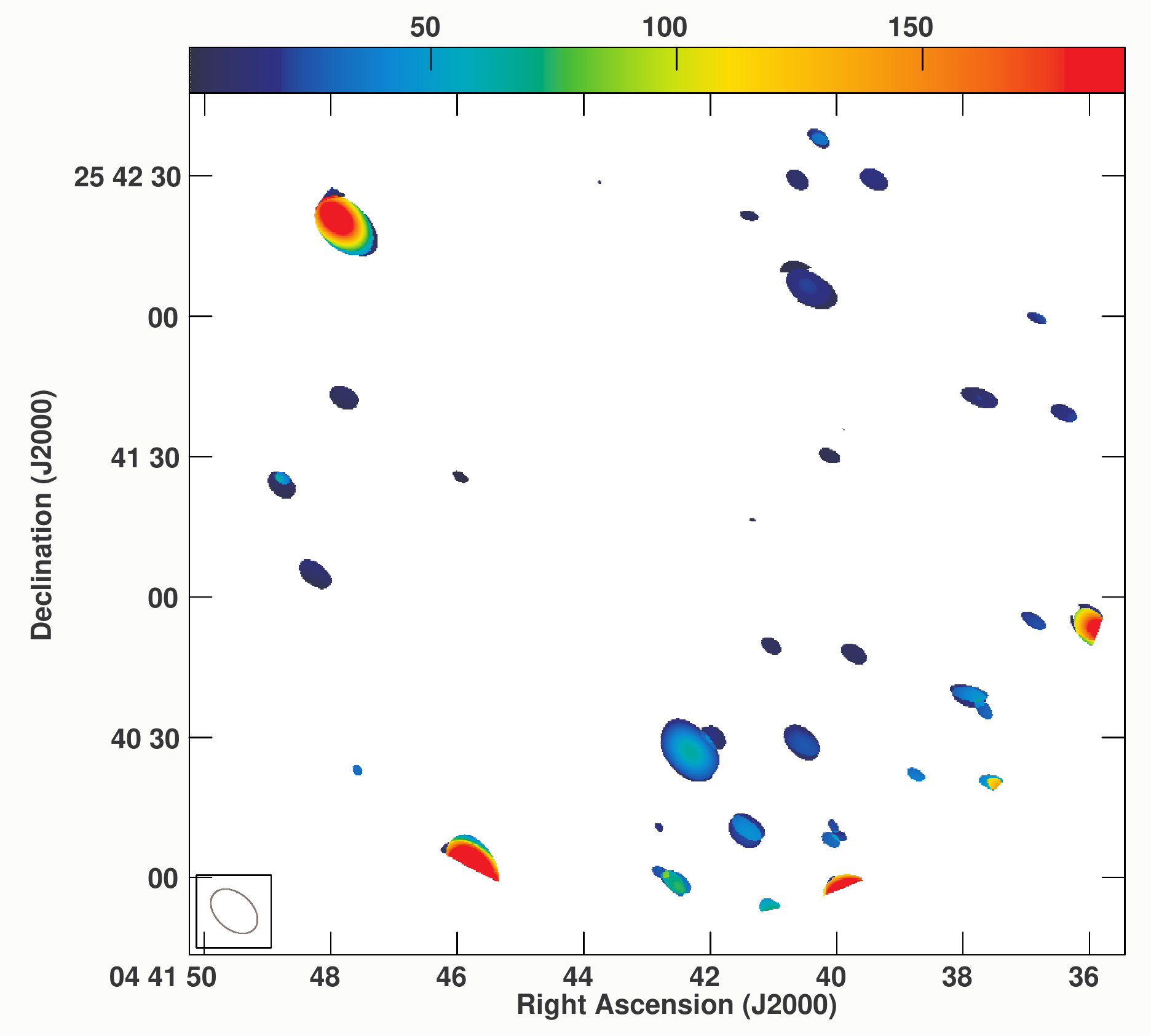}\\
  \caption{Integrated spectra of NH$_{3}$(J,K = 1,1) 23.69 GHz line towards the TMC-1 core. RMS noise of the spectra is $\sim$ 0.77 Jy. (b) Integrated intensity map of  NH$_{3}$(J,K = 1,1) 23.69 GHz line emission towards the TMC-1 core at a angular resolution of 11.43$''$ $\times$7.69$''$.}
 \label{fig:fig13}
\end{figure*}

\section{Observation of TMC-1C, L1544, and TMC-1 cores from VLA C and CNB configurations with  CCS and NH$_{3}$ :}\label{three cores}

We have also studied the cores TMC-1C, L1544, and TMC-1 with VLA in C and CNB configurations. The physical and kinematical properties of TMC-1C have already been discussed in the previous sections. \citet{2018ApJ...864...82D,2019ApJ...879...88D} did the radiative transfer modeling of the  TMC-1 core with CCS(4$_{3}$-3$_{2}$), HC$_{3}$N(J=5-4) main and hyperfine lines for the inner region and with $^{13}$CO(J=1-0), $^{18}$CO(J=1-0) lines for the diffuse outer region. From their analysis, they found the inward motion of this core. They discussed different possibilities of this inward motion, e.g., pure gravitational contraction (if not supported by magnetic field) or oscillation of the filament. \citet{2019PASJ...71..117N} measured the line of sight magnetic field $\sim$117  $\pm$ 21 $\mu$G with CCS 45.4 GHz. From their analysis, TMC-1  is magnetically supercritical. From the studies of \citet{1992ApJ...392..551S} it was found that in TMC-1, the ratio of N$_{NH_{3}}$ and N$_{CCS}$ is $\sim$ 2.9. Different organic molecules in this core have been studied by \citet{2018ApJ...854..116S}.\\
Like TMC-1C and TMC-1, L1544 is also one pre-stellar core in the Taurus molecular cloud. Previously, \citet{1999ApJ...518L..41O} studied this core with CCS radical and found that CCS emission has a central hole of 7,500 AU. The reason for this is the depletion of CCS radical with chemical evolution of the core, as predicted by \citet{2003ApJ...593..906A}. From the studies of \citet{1992ApJ...392..551S}, it was found that in L1544, the ratio of N$_{NH_{3}}$ and N$_{CCS}$ is $\sim$ 15. It indicates that it is an evolved core.\\

Observational data for this analysis were taken from the VLA archival data center (12A-402). The observations were carried out in 2012 from March to May. Cores are studied with CCS 2$_{1}$- 1$_{0}$ and NH$_{3}$(1,1) at 22.344, 23.694 GHz respectively. TMC-1C and TMC-1 were observed with VLA in C configuration, whereas L1544 in CNB configuration. At this frequency, angular resolutions of VLA C and CNB configurations are  0.95 and 0.28 arcseconds, respectively. RMS noise has been achieved $\sim$ 3 mJy beam$^{-1}$ for almost three hours of on-source observation. Spectral resolution for these observations is $\sim$3.9 kHz ($\sim$ 0.05 km sec$^{-1})$.\\

Initial flagging and calibration were done manually in AIPS (Astronomical Image Processing System; version 31DEC19). To remove any weak continuum, we used the task \textbf{UVLIN} to subtract the continuum by using the line-free channels. Doppler
correction was performed using the task \textbf{CVEL}. For imaging, we used the multiscaling with the task \textbf{IMAGR} to pick up different structures. Different scales that we have used are mentioned in table \ref{tab:tab1}.\\
\\
\\

\begin{table}
\caption{Different used scales for TMC-1C, L1544, and TMC-1.}
\begin{center}

\begin{tikzpicture}
\draw (0,0) -- (8,0);
\end{tikzpicture}
\begin{tabular}{llll}

 	&  TMC-1C	&  &  \\
CCS & 4.45$''$ $\times$ 3.74$''$ & 11.90$''$ $\times$ 8.90$''$ & 19.05$''$ $\times$ 17.77$''$\\
NH$_{3}$ & 4.42$''$ $\times$ 3.71$''$ & 11.67$''$ $\times$ 8.80$''$ & 18.10$''$ $\times$ 17.11$''$\\
\\
 	&  L1544	&  &  \\
CCS & 4.15$''$ $\times$ 3.51$''$ & 7.23$''$ $\times$ 5.99$''$ & 11.96$''$ $\times$ 9.77$''$\\
NH$_{3}$ & 3.03$''$ $\times$ 2.41$''$ & 5.26$''$ $\times$ 4.27$''$ & 8.63$''$ $\times$7.23$''$\\  
\\
 	&  TMC-1	&  &  \\
CCS & 4.80$''$ $\times$ 4.21$''$ & 7.77$''$ $\times$ 6.25$''$ & 11.87$''$ $\times$ 8.04$''$\\
NH$_{3}$	& 4.73$''$ $\times$ 4.14$''$ & 7.64$''$ $\times$ 6.03$''$ & 11.43$''$ $\times$7.69$''$\\

\end{tabular}
\end{center}
\begin{tikzpicture}
\draw (0,0) -- (8,0);
\end{tikzpicture}

\label{tab:tab1}

\end{table}


CCS (2$_{1}$-1$_{0}$) and NH$_{3}$ (J,K = 1,1) main hyperfine lines are detected in almost all these three cores. Their spectra are shown in the left panels of the figures \ref{fig:fig8}, \ref{fig:fig9}, \ref{fig:fig10}, \ref{fig:fig11}, \ref{fig:fig12} and \ref{fig:fig13} whereas their integrated emissions are shown in the right panels of the same figures. The peak line flux densities of CCS lines are detected in these observations are approximately at 3.8, 0.32, and 7.46 Jy for the TMC-1C, L1544, and TMC-1 cores respectively. Similarly, the peak line flux densities of NH$_{3}$(J,K = 1,1) lines are detected in these observations are approximately at 3.8, 0.15, and 5.32 Jy for TMC-1C, L1544, and TMC-1 cores, respectively.\\
We have found that structures of CCS and NH$_{3}$ molecules are clumpy, and most of the signal is resolved out in these configurations, as well as only the main hyperfine line of ammonia is detected due to sensitivity issues.
Spatial incoherence of the thioxoethenylidene (CCS) and ammonia (NH$_{3}$) is also observed from the integrated intensity maps in these cores. A possible argument is that these molecules trace different environments of the cores.\\
We have compared the integrated fluxes with the previous single-dish telescope. As in our analysis, we have not successfully recovered all the hyperfine lines of ammonia; the comparison of fluxes is only made for CCS spectra. CCS integrated fluxes in TMC-1C and TMC-1 observed with VLA in C configuration are 0.55 Jy km sec$^{-1}$ and 1.95 Jy km sec$^{-1}$ respectively. Whereas in L1544, recovered integrated flux recovered is 0.11 Jy km sec$^{-1}$. CCS integrated flux densities observed in TMC-1C, L1544 and TMC-1 with Nobeyama 45 meter single-dish telescope are 1.932, 1.288 and 2.94 Jy km sec$^{-1}$ respectively \citep{1992ApJ...392..551S}. As noted earlier, at this 22.3 GHz frequency, the FWHM of the Nobeyama 45-meter telescope is $\sim$ 75$''$. For a one-to-one comparison, we cut the image cube into a circle whose center is the pointing center and diameter is $\sim$ 75$''$. Then we have obtained the integrated fluxes 0.027, 0.039, and 0.046 Jy km sec$^{-1}$ for TMC-1C, L1544, and TMC-1, respectively. Thus, from the interferometric observations, we recover only $\sim$ 1.4\%, 3.0\% and 1.5\% for TMC-1C, L1544 and TMC-1 respectively. Earlier, \cite{2011ApJ...739L...4R} studied three cores, TMC-1C, TMC-1, and L1521B, with VLA in DNC configuration using the CCS 22.3 GHz transition. In their study, they recovered only 1- 13$\%$ of the total flux observed by single dish telescope. Similarly, \cite{2007A&A...470..221C} studied the L1544 core with VLA in D configuration using the ammonia molecule and found a significant fraction($\sim$70$\%$) of flux is resolved out in the interferometric observation. \


\section{Conclusions}\label{conclusions}

We have studied the TMC-1C, L1544, and TMC-1 cores with VLA in different configurations using the thioxoethenylidene (CCS) radical and ammonia (NH${_3}$) molecule. Our observations and analysis reveal the following:\\

(i) We observe the infall of the central part of the TMC-1C core through CCS radical. It indicates that the TMC-1C core is on the verge of star formation.\\

(ii) We calculate the turbulence inside the TMC-1C core and are found to be subsonic. It might create a suitable environment for the star formation.\\

(iii) We perform the virial analysis in the TMC-1C core and find that the virial parameter $\alpha$ is $<$ 1. It suggests that thermal and non-thermal motions cannot prevent the collapse.\\

(iv) We compare the integrated intensity of CCS with the previous single-dish observation and find that a large amount of emission is resolved in interferometric observation. Spatial offset between CCS and dust 1200$\mu$m integrated emissions is also observed in TMC-1C core. It may indicate that CCS was present in copious amounts in the less dense environments at the early stage of the evolution. However, during the evolution of the core, it is depleted and freezes out into cold dust grains.\\

(v) We find the spatial incoherence of CCS radical and NH${_3}$ molecule, probing the different environments of the cores.\\

This work demonstrates the possibility of studying the initial conditions of the prestellar cores in high spatial resolution interferometric observation. Future deep sensitive observations of CCS and NH$_{3}$ in these prestellar cores will enable us to study the magnetic field (through CCS) and kinetic temperature (through NH$_{3}$) in the extremely dense nucleus of the cores.\\

\section*{Acknowledgements}
We thank the anonymous reviewer for the valuable comments that have helped to improve this paper. A.K. would like to thank DST-INSPIRE (IF160553) for the fellowship. We are grateful to Dr. Nirupam Roy and Dr. Thushara G.S. Pillai for their useful discussion.\\

\section*{Data Availability}
The data products from this study will be shared on reasonable request to the corresponding author.

\bibliographystyle{mnras}

\bibliography{mnras}

\appendix 

\end{document}